\documentclass[%
 reprint,
 amsmath,amssymb,
 aps,
]{revtex4-2}

\usepackage{graphicx}
\usepackage{dcolumn}
\usepackage{bm}
\usepackage{upgreek}
\usepackage{mathtools}
\usepackage{multirow}
\usepackage{xcolor}
\usepackage{hyperref}
\usepackage[normalem]{ulem}
\usepackage{placeins}

\hypersetup{
  colorlinks   = true, 
  urlcolor     = black, 
  linkcolor    = blue, 
  citecolor   = magenta 
}
\def\RPlus{\rule[0.165em]{.5em}{.165em}\hspace{-.33em}\rule[0em]{.165em}{.5em}\,}
\usepackage{xr}
\makeatletter

\newcommand*{\addFileDependency}[1]{
\typeout{(#1)}
\@addtofilelist{#1}
\IfFileExists{#1}{}{\typeout{No file #1.}}
}\makeatother

\begin{document}
\title{\textcolor{black}{StrainTensorNet}: Predicting crystal structure elastic properties using \\ SE(3)-equivariant graph neural networks}
\author{Teerachote Pakornchote}
\author{Annop Ektarawong}
\author{Thiparat Chotibut}
\email[Correspondence to: ]{thiparatc@gmail.com}
\affiliation{%
 Chula Intelligent and Complex Systems, Department of Physics, Faculty of Science, Chulalongkorn University, Bangkok, Thailand, 10330
}

\date{\today}
\begin{abstract}
Accurately predicting the elastic properties of crystalline solids is vital for computational materials science. However, traditional atomistic scale \textit{ab initio} approaches are computationally intensive, especially for studying complex materials with a large number of atoms in a unit cell. We introduce a novel data-driven approach to efficiently predict the elastic properties of crystal structures using SE(3)-equivariant graph neural networks (GNNs). This approach yields important scalar elastic moduli with the accuracy comparable to recent data-driven studies. Importantly, our symmetry-aware GNNs model also enables the prediction of the strain energy density (SED) and the associated elastic constants, the fundamental tensorial quantities that are significantly influenced by a material's crystallographic group.  The model consistently distinguishes independent elements of SED tensors, in accordance with the symmetry of the crystal structures. Finally, our deep learning model possesses meaningful latent features, offering an interpretable prediction of the elastic properties.

\begin{description}
\item[Keywords]
strain energy tensor, elastic constants, crystallographic group, \\ equivariant neural networks, density functional theory
\item[Open data] The dataset used in this work has been made available at \\ \url{https://github.com/trachote/predict\_elastic\_tensor}
\end{description}
\end{abstract}

\maketitle

\section{Introduction}
 The elastic properties of crystalline solids such as elastic constants, bulk modulus, and shear modulus are important macroscopic quantities that determine materials' mechanical characteristics. Computational study of which can provide theoretical guidelines to understand various phenomena in solid materials, e.g., mechanical stability of crystal structures \cite{Born1988, Mouhat2014},  pressure-driven deformation and phase transition of materials \cite{Macdonald1969, Brazhkin2007}, the response of materials to sound wave propagation \cite{Yamaguchi1997, Reichmann2000}, and the hardness of materials \cite{Teter1998, Chen2011, Tian2012}, to name a few. From an atomistic scale description, an \textit{ab initio} approach based on density functional theory has been employed to investigate the macroscopic elastic properties of materials, yielding, for example, elastic constants and elastic moduli that are in agreement with experimental results \cite{Nielsen1985, Labeguerie2010, Wang2010, Ektarawong2016}. 
 
Three atomistic scale computational methods are usually adopted to calculate the elastic constants of crystal structures: an energy-based method \cite{Stadler1996}, a stress-strain method \cite{Nielsen1985, Page2002}, and a calculation from group velocity \cite{Sarasamak2010, Chernozatonskii2011, Pakornchote2021}. The energy-based and the stress-strain methods are more standard, and derive the elastic constants using respectively an energy and a stress tensor acquired from an \textit{ab initio} calculation. 
Although both methods yield a comparable prediction of elastic constants, the energy-based method is computationally less efficient, involving a larger number of plane-wave basis set and $k$-point meshes so that the predicted elastic constants converge to reliable values~\cite{Caro2012}. 
Indeed, the more efficient stress-strain method is commonly employed to determine elastic constants by established softwares such as VASP, Quantum Espresso and CASTEP~\cite{Kresse1996, Giannozzi2009_qe, Clark2005}. The stress-strain method is also utilized to create a database for elastic properties of inorganic compounds \cite{deJong2015, Jain2013}.

However, atomistic scale simulation can be computationally prohibitive, especially when the number of atoms in a unit cell grows large. Such constraint limits an {\it ab initio} method's capability to investigate more complex crystalline solids. On the other hand, advances in machine learning bring about alternative approaches to computational materials science. This data-driven paradigm can reasonably predict the elastic properties of crystal structures, provided sufficient training data from an {\it ab initio} method \cite{Mazhnik2020, Zhao2020}. Even when the number of atoms in a unit cell is large, such approach can efficiently predict the bulk and shear moduli of complex materials such as alloys \cite{Levamaki2022, Linton2023}. Applying machine learning together with 
{\it ab initio} calculations is also potentially useful in searching for novel superhard materials \cite{Avery2019, Chen2021}.

Machine learning models based on graph neural networks (GNNs) have received increasing attention in studying solid materials and molecules. With GNNs, it is natural to encode atomistic scaled descriptions of solids into a computational framework; atomic locations and the associated atomic attributes can be directly embedded into node attributes of a graph, whereas pairwise interactions among atoms can be encoded into the edge attributes. 
Efficient GNNs training procedures have also been proposed \cite{Kearnes2016, Gilmer2017, Schutt2018}, enabling GNNs to learn the representation of a complex relationship between the input and its associated prediction \cite{Dym2021}. These neural networks can also be endowed with translation-rotation equivariant property, so that input atomic locations of molecules (or point clouds) in $\mathbb{R}^3$ that differ only in their orientation or their centroid can be identified. Enforcing an SE(3) equivariant property helps the networks to extract a more compact (translation and rotation independent) relationship between the inputs and their predictions. Due to these appealing features, variations of SE(3) equivariant GNNs have been developed to study materials science \footnote{Tensor Field Network (TFN), L1Net, GemNet, and EGNN, have exploited this computational architecture to make better data-driven predictions on the properties of chemical compounds \cite{Thomas2018, Miller2020, Klicpera2021, Garcia2021}. The SE(3)-Transformers model based on TFN with an attention mechanism also demonstrates a good predictive performance on the standard QM9 dataset \cite{Fuchs2020}. In addition, translation-rotation equivariance concept has also been incorporated into generative modeling, in an attempt to search for novel molecules and crystal structures with desirable properties \cite{Xu2022, Xie2022}.}. 

In this work,  we adopt a data-driven approach using graph neural networks (GNNs) to predict the elastic properties of crystal structures. Accounting for the symmetry of crystalline solids, our SE(3) equivariant GNNs take as an input atomistic descriptions of strained materials, such as atomic locations and their associated atomic attributes, and predict the strain energy density (SED) of the materials. 
The prediction of the SED, which is the energy stored in a crystal structure when distorted by either tensile or compressive strains in particular directions, can be obtained efficiently and relatively accurately, given sufficient training data. The model thus provides an alternative approach to the standard {\it ab initio} energy-based prediction method. 

After the SED is computed, we can then calculate the elastic constants and construct the elastic tensor with a simple analytical expression, see Sec.~\ref{sec: SET}.  Other macroscopic (scalar) elastic properties including bulk modulus, shear modulus, Young's modulus, and Poisson's ratio immediately follow from the elastic constants.  Sec.~\ref{sec: results} reports our model prediction results of these elastic properties. The prediction performances of the scalar elastic properties are comparable to those of the recent data-driven work \cite{Mazhnik2020, Zhao2020}. Importantly, however, the data-driven prediction of the SED and the associated elastic constants, which are fundamental tensorial quantities that depend on the crystallographic group, are first reported here. The trained model consistently reveal the independent components of strain energy tensors dictated by the symmetry of the crystal structures, as evident by the degeneracy structure of the distance metrics computed in Sec. \ref{sec:pred_result}.  Our symmetry-aware GNNs architecture as well as the dataset used to train the model are provided in Sec.~\ref{sec: ML}. Also, as opposed to a black-box deep learning prediction, we show in Sec.~\ref{sec: ML-latent} that the latent features (data representations) that the model learned are physically meaningful, rendering our GNNs explainable. 
We conclude this work in Sec.~\ref{sec: conclusion}, and provide supplementary results and necessary derivations in the Appendices. 

\section{Linear Elasticity Background}\label{sec: elasticity}
\subsection{The strain energy tensor and the elastic tensor}\label{sec: SET}
Assuming linear elastic deformation and no external stress, the strain energy density (SED) of isotropic materials can be expressed as \cite{Slaughter2002}
\begin{equation} \label{eq:u123456}
    U(\epsilon_1, \epsilon_2, ..., \epsilon_6) = \frac{1}{2}\sum_{i,j=1}^{6}C_{ij}\epsilon_i\epsilon_j,
\end{equation}
where $C_{ij}$ is an {\it elastic constant} and $\epsilon_i$ is a strain component $i \in \{1,2, \dots,6 \}$ in Voigt notation. \textcolor{black}{With the goal of obtaining the elastic constant, it suffices to focus on the SED tensor of rank 2, which we now show.  Denote  $U_{ij} \equiv U(\epsilon_i,\epsilon_j)$ for $i \neq j$ and $U_{ii} \equiv U(\epsilon_i)$, which are the energy stored in a distorted crystal structure when the lattice is strained by at most two strain components (two Voigt indices).} In this study, we consider $U_{ij}$ in units of eV per atom.
Since the elastic constant tensor is symmetric, the SED tensor is also symmetric and can be expressed as
\begin{equation} \label{eq:uij}
\begin{aligned}
    U_{ij} &= C_{ij}\epsilon_i\epsilon_j + \frac{1}{2}\left(C_{ii}\epsilon_i^2+C_{jj}\epsilon_j^2\right)  \text{\ for\ } i \neq j \text{ and,} \\
    U_{ii} &= \frac{1}{2}C_{ii}\epsilon_i^2.
\end{aligned}
\end{equation}
Thus, the elastic constants can then be analytically obtained as a function of the SED:
\begin{equation} \label{eq:cij}
    C_{ij}=\frac{1}{\epsilon_i\epsilon_j}\left[(1+\delta_{ij})U_{ij}-\left(1-\delta_{ij}\right)\left(U_{ii}+U_{jj}\right)\right],
\end{equation}
where $\delta_{ij}$ is the Kronecker delta.

Because the SED tensor \textcolor{black}{of interest} and the elastic tensor are symmetric tensors of rank 2 with $6 \times 6$ components, \textcolor{black}{so} there are $(6+1)6/2 = 21$ independent components. In the matrix form, we denote the SED tensor as 
\begin{equation} \label{eq:u_general}
    \mathbf{U} = 
    \begin{bmatrix}
    U_{11}&U_{12}&U_{13}&U_{14}&U_{15}&U_{16}\\[3pt]
     &U_{22}&U_{23}&U_{24}&U_{25}&U_{26}\\[3pt]
     & &U_{33}&U_{34}&U_{35}&U_{36}\\[3pt]
     & & &U_{44}&U_{45}&U_{46}\\[3pt]
     & & & &U_{55}&U_{56}\\[3pt]
     & & & & &U_{66}
    \end{bmatrix}
    ,
\end{equation}
where only the upper triangular elements are shown (the lower diagonal elements are left blank for brevity). For crystalline solids, however, the upper triangular elements are not completely independent, depending on the symmetry of the crystal structure. For instance, the SED tensor of a cubic lattice is 
\begin{equation} \label{eq:u_cubic}
    \mathbf{U}_{cubic} = 
    \begin{bmatrix}
    U_{11}&U_{12}&U_{12}&U_{14}&U_{14}&U_{14}\\[3pt]
     &U_{11}&U_{12}&U_{14}&U_{14}&U_{14}\\[3pt]
     & &U_{11}&U_{14}&U_{14}&U_{14}\\[3pt]
     & & &U_{44}&U_{45}&U_{45}\\[3pt]
     & & & &U_{44}&U_{45}\\[3pt]
     & & & & &U_{44}
    \end{bmatrix}
    .
\end{equation}

Note that some components of the elastic constants $C_{ij}$ can be zero in many crystal structures, e.g. a cubic lattice has 9 zeros out of 21 independent components. However, due to the property of SED from Eq. \eqref{eq:uij}, $U_{ij}$ is never 0. For the purpose of machine learning regression, working with SED helps avoid zero inflation problems in the training dataset. Fig.~\ref{fig:data_distributions} shows the distributions of $C_{ij}$ and $U_{ij}$, which cover different ranges depending on the indices. Importantly, the distribution of elastic constants can concentrate around zero, but this is not the case for the SED. As an illustrative example of how the SED can avoid a zero inflation problem, consider the elastic constants of a diamond cubic crystal structure. For diamond, $C_{11}$, $C_{12}$, $C_{14}$, $C_{44}$, and $C_{45}$ are 1054, 126, 0, 562, and 0 GPa, respectively. Then, Eq.~\eqref{eq:uij} gives $U_{11}$, $U_{12}$, $U_{14}$, $U_{44}$, and $U_{45}$ subjected to 2\% strain to be 7.506, 16.807, 11.509, 4.002, and 8.005 meV/atom, respectively. Note that, in this study, we consider $U_{ij}$ in the unit of eV per atom instead of per volume. The magnitude of $U_{12}$ is the largest as it is the sum of $C_{12}$ and $C_{11}$, while $U_{14}$ and $U_{45}$ are smaller (but non-zero) because $C_{14}$ and $C_{45}$ are zero. 

\subsection{Strain operations}\label{sec: strain operation}

The unit cell of a crystal can be compactly parametrized by a lattice matrix $\mathbf{L}=[\mathbf{a}\;\mathbf{b}\;\mathbf{c}]$, where $\mathbf{a}$, $\mathbf{b}$, and $\mathbf{c}$ are vectors of lattice parameters in the Cartesian coordinate. If the system is applied by strain, the lattice matrix will be deformed by
\begin{equation}
\begin{aligned} \label{eq:lattice}
   \mathbf{L}^{\prime}(\epsilon_1, \epsilon_2, ..., \epsilon_6)  &= \bm{\upvarepsilon}_I\mathbf{L},
\end{aligned} 
\end{equation}
where $\bm{\upvarepsilon}_I$ is the strain matrix
\begin{equation}\label{eq:strain_matrix}
    \bm{\upvarepsilon}_I = 
    \begin{bmatrix}
    \epsilon_1&\frac{\epsilon_6}{2}&\frac{\epsilon_5}{2}\\[3pt]
    \frac{\epsilon_6}{2}&\epsilon_2&\frac{\epsilon_4}{2}\\[3pt]
    \frac{\epsilon_5}{2}&\frac{\epsilon_4}{2}&\epsilon_3\\
    \end{bmatrix} + \mathbf{I}.
\end{equation}
In this work, we assume the crystal structures will be deformed by at most two strain components, so that $\mathbf{L}^{\prime}=\mathbf{L}^{\prime}(\epsilon_i, \epsilon_j).$

Due to an applied strain, the atomic coordinates $\mathbf{r}$ must also be transformed accordingly as 
\begin{equation} \label{eq:strained_coord}
    \mathbf{r}' = \bm{\upvarepsilon}_I\mathbf{L}\mathbf{r}_f,
\end{equation}
where $\mathbf{r}_f$ is a fractional coordinate of an atom in an unstrained lattice. Noting that $\mathbf{r}_f$ typically shifts from its equilibrium value when the strain is applied (say, in density function theory or in experiments); however, the atomic relaxation is neglected for simplicity.

\begin{table*}[ht]
\caption{\label{tab:crystal_system}The minimal set of $ij$ strained component of each crystal system used to generate an initial crystal graph input}
\begin{ruledtabular}
\begin{tabular}{lcc}
Crystal systems&\# of training indices&Training indices\\
\hline
cubic&6&11, 12, 14, 15, 44, 45\\
hexagonal&13&11, 12, 13, 14, 15, 16, 33, 34, 36, 44, 45, 46, 66\\
tetragonal&14&11, 12, 13, 14, 15, 16, 26, 33, 34, 36, 44, 45, 46, 66\\
trigonal&16&11, 12, 13, 14, 15, 16, 24, 25, 33, 34, 36, 44, 45, 46, 66\\
orthorhombic&21&all\\
monoclinic&21&all\\
triclinic&21&all\\
\end{tabular}
\end{ruledtabular}
\end{table*}

\section{Machine Learning with SE(3)-Equivariant Graph Neural Networks}\label{sec: ML}
\subsection{Crystal graphs}
\begin{figure}[t]
\includegraphics[width=.48\textwidth]{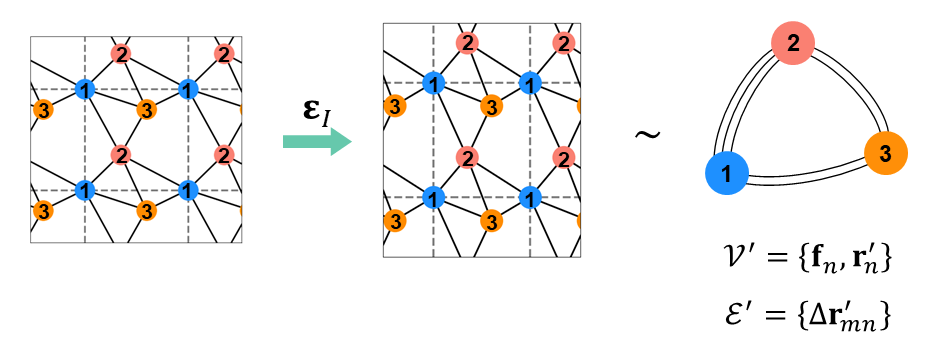}
\caption{\label{fig:crystal_graph} A 2-dimensional lattice and its {\it strained} crystal graph embedding. Grey dashed lines are periodic boundaries. The whole structure can be spanned by the 3 atoms in 3 different bases in the unit cell, represented by 3 different nodes. Black lines are (undirected) edges connecting two neighbor nodes.  
The strained lattice is represented as a multigraph on the right, which is the crystal graph input into our SE(3)-equivariant GNN building block. Note that to describe the crystal graph for the unit cell and the atoms on its boundary, only three nodes are required. The multi-edges of each pair of nodes are distinguished by different crystallographic directions.
}
\end{figure}

A crystal structure can be represented as a multigraph whose nodes and edges, respectively, encode atoms and their pairwise connections. 
A pair of nodes describing an atom of type $m$ and an atom of type $n$ can be connected by multiple edges, encapsulating the interactions between the atom of type $m$ in a unit cell and atoms of type $n$ in the unit cell as well as in other periodic cells of consideration, see Fig \ref{fig:crystal_graph}. An atom of type $m$ can interact with an atom of the same type in periodic cells if the interaction range of consideration is large enough, represented by self-loops in the node $m$ of the crystal graph. Each atom's location and its atomic features (e.g., atomic mass, electronegativity, polarizability, atomic radius, and etc.) in a unit cell is accounted for by the attributes of a single node. 

More formally, a crystal graph $\mathcal{G} = (\mathcal{V}, \mathcal{E}$) consists of a set of nodes (vertices) $\mathcal{V}$ and a set of edges $\mathcal{E}$ defined as
\begin{eqnarray*}
    &\mathcal{V} = \{(\mathbf{f}_n, \mathbf{r}_n) \; | \; \mathbf{f}_n \in \mathbb{R}^{M}, \; \mathbf{r}_n = \mathbf{L} \mathbf{r}_{f_n} \in \mathbb{R}^{3} \}   , \\
    &\mathcal{E} = \{\Delta\mathbf{r}_{mn}^{(\mathbf{T})} \; | \; \Delta\mathbf{r}_{mn}^{(\mathbf{T})} = \mathbf{r}_m - \mathbf{r}_n + \mathbf{T}; \ \mathbf{r}_m, \mathbf{r}_n \in \mathbb{R}^{3} \} ,
\end{eqnarray*}
where $m$ and $n$ are indices of the two atoms in the unit cell, $\mathbf{f}_n$ is a vectorized atomic feature of an atom of type $n$, $M$ is the number of atomic features, $\mathbf{r}_{f_n}$ is the fractional coordinate of an atom of type $n$, and $\mathbf{T}$ is a translation vector within the space of interest.
Each node embeds two kinds of information: atomic attributes $\mathbf{f}$ and atomic positions $\mathbf{r}$ in the unit cell. From this definition, the unit cell and its periodic images is compactly represented as a multigraph; only a finite number of nodes describing the atoms in the unit cell is required, and the number of edges joining the two nodes grows linearly with the number of periodic images of the atoms in the structure of interest (see Fig.~\ref{fig:crystal_graph}).


A crystal graph describing the whole crystal structure will require infinitely many edges information. In practice, one typically embeds only partial information of the structure using the description of the unit cell and a finite number of edges encoding interactions with neighbor atoms \cite{Xie2018}. One method to construct the edges is to connect a target atom to other atoms within a finite radius; however, still, this method generates an excessive number of edges, which is computationally infeasible for GNNs machine learning. Alternatively, some algorithms determine the connection between atoms using, e.g., the Voronoi tessellation, generating a moderate number of edges \cite{Zimmermann2017, Pan2021}. However, for layered materials, using Voronoi tessellation causes the edges connecting atoms between layers to be absent. These interlayer connections are crucial in differentiating the structure strained in out-of-plane direction from the structures strained in other directions. In this work, we define an edge between a pair of atoms $\Delta \mathbf{r}$ (dropping the translation vector superscript and the atom indices subscript for brevity) to be non-zero only if each component $\Delta r_{\beta}$ where ${\beta} \in \{x,y,z\}$ satisfies the following spatial cutoff criterion: 
$|\Delta r_{\beta}| \leq \min(|a_{\beta}|+|b_{\beta}|+|c_{\beta}|, s)+\delta$ where $a_{\beta}$, $b_{\beta}$, and $c_{\beta}$ are the $\beta$ components of the corresponding lattice parameters constituting the lattice matrix $\mathbf{L}$, $s$ is a cutoff distance, and $\delta$ is a small cutoff extension of our choice. This can reduce the number of edges from the hard radial cutoff criterion and ensures the connections of atoms between layers for layered materials by an appropriate choice of $s$ and $\delta$.

\subsection{Rotational and translational equivariance}

If two crystal graph inputs constructed from an identical crystal graph strained by two different sets of strained components $(\epsilon_i,\epsilon_j)$ and $(\epsilon_k, \epsilon_l)$ yielding the new vertices $\mathcal{V'} = \{ (\mathbf{f}_n, \mathbf{r'}_n) \}$ and $\mathcal{V''} = \{ (\mathbf{f}_n, \mathbf{r''}_n) \}$, respectively, are equivalent up to a 3D rotation and a translation, then we expect our machine learning model to predict the same SED. This symmetry-aware property can be implemented with  geometric deep learning models \cite{Bronstein2021}. We will equip a latent representation of our model with this symmetry-aware (equivariant) property, which is defined \textcolor{black}{as follows}.

A latent representation $\phi: \mathcal{V} \rightarrow \mathcal{Y} $ is {\it equivariant} if for each transformation $T_g:  \mathcal{V} \rightarrow \mathcal{V} $ with $g$ being an element of an abstract group $G$, there exists a transformation $S_g:  \mathcal{Y} \rightarrow \mathcal{Y} $ such that the following condition holds: 
$$
S_g[\phi(v)]=\phi\left(T_g[v]\right),
$$
for all $g \in G, v \in \mathcal{V}.$ 
In our case, the group $G$ of interest is SE(3), providing the latent representation with a 3D rotation and translation equivariant property. This latent feature $\phi$ can be achieved with SE(3)-equivariant GNNs, known as Tensor Field Network (TFN) \cite{Thomas2018}, and with a TFN with an appropriate attention mechanism, known as SE(3)-transformers \cite{Fuchs2020}, see more detailed recipes of these two models in Appendix~\ref{sec: SE(3)-transformers}. 

The key idea that enables SE(3)-equivariant property of these two GNNs is that its message passing kernel is constructed from translational invariant spherical harmonic bases, with the structure-preserving transformation $S_g$ given by the Wigner-D matrices (see Appendix~\ref{sec: SE(3)-transformers}). Under a 3D rotation, each spherical harmonic $J$ basis function of the kernel transforms according to 
$$
Y_J\left(\mathbf{R}_g^{-1}\frac{\Delta\mathbf{r}}{||\Delta\mathbf{r} ||}\right)=\mathbf{D}_J^\ast(g) Y_J\left(\frac{\Delta\mathbf{r}}{||\Delta\mathbf{r} ||}\right),
$$
where $\mathbf{D}_J(g)$ is the $J^{th}$ Wigner-D matrix, and $\mathbf{R}_g$ is a rotation matrix associated with $g \in \mathrm{SO}(3)$, making the learned latent representation $\phi$ equivariant \cite{Thomas2018}. The multi-head attention mechanism in SE(3)-transformers also induces an equivariant property on the product rule between each key and value pairs, rendering the whole message passing algorithm equivariant under SE(3) \cite{Fuchs2020}. In this work, we use SE(3)-transformers to first build a compressed representation of the orientations of a strained crystal structure, then such learned latent representation will be passed to other networks with a high expressive power to perform the prediction (regression) task of the SED. The following section contains the entire framework for the SED prediction.

\subsection{Model architecture and training procedure}\label{subsec: model and training}

\begin{figure*}[t]
\includegraphics[width=.99\textwidth]{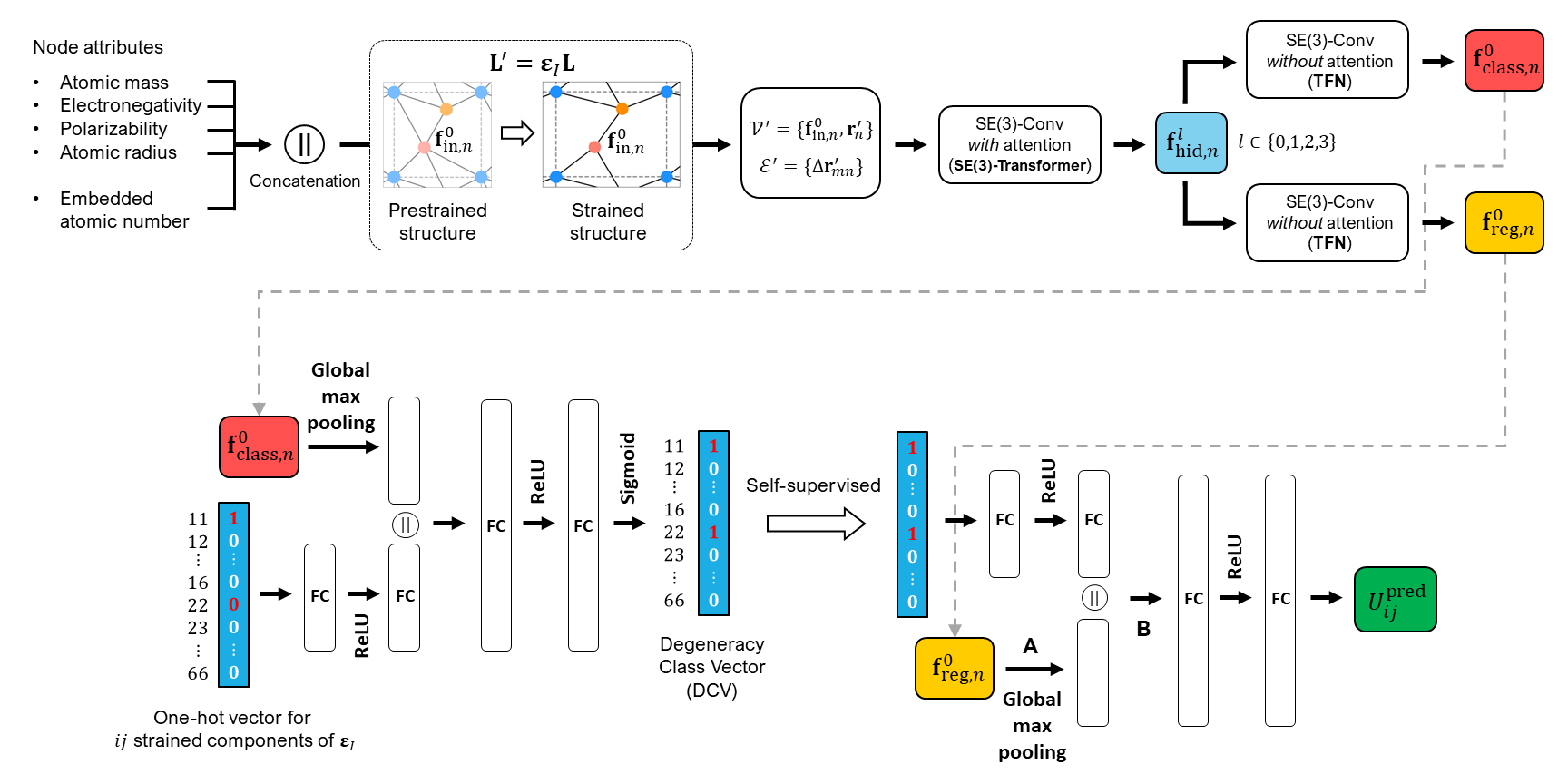}
\caption{\label{fig:model} With a strained crystal graph and the strained component one-hot vector as the input, \textcolor{black}{StrainTensorNet} employs a self-supervised approach combining both classification and regression networks to predict both the degeneracy class vector and the $U_{ij}$. The $\mathbf{f}_{\mathrm{in},n}^0 \equiv \mathbf{f}_{n}$ denotes the input feature of the $n^{\textrm{th}}$ node, where a superscript, 0, indicates the input into the $l=0$ channel of the spherical harmonics in SE(3)-Transformer, which is a channel whose feature is invariant under rotation (transforms like a scalar quantity). $\Delta \mathbf{r}'_{mn}$ is the edge connecting a target node $m$ and a neighbor node $n$. The degeneracy class vector from the classification network will be concatenated with the latent representation of the strained crystal graph ($\mathbf{f}^0_{\mathrm{reg},n}$) for the final regression network to predict the SED. }
\end{figure*}

We now describe our GNNs-based model that predicts the SED tensor.
The architecture of our model, which we term {\it \textcolor{black}{StrainTensorNet}}, is illustrated in Fig.~\ref{fig:model}. The model is designed to receive two types of input: a crystal graph representing a {\it strained} crystal structure, and a one-hot vector of dimension $6(6+1)/2 = 21$ \footnote{Due to the symmetry of SED, we identify only an upper triangular element} indicating the {\it strained components}, with a value 1 in the $ij$ component that the strain operation is applied on and with a value 0 otherwise. 

\begin{figure*}[ht]
\includegraphics[width=1.\textwidth]{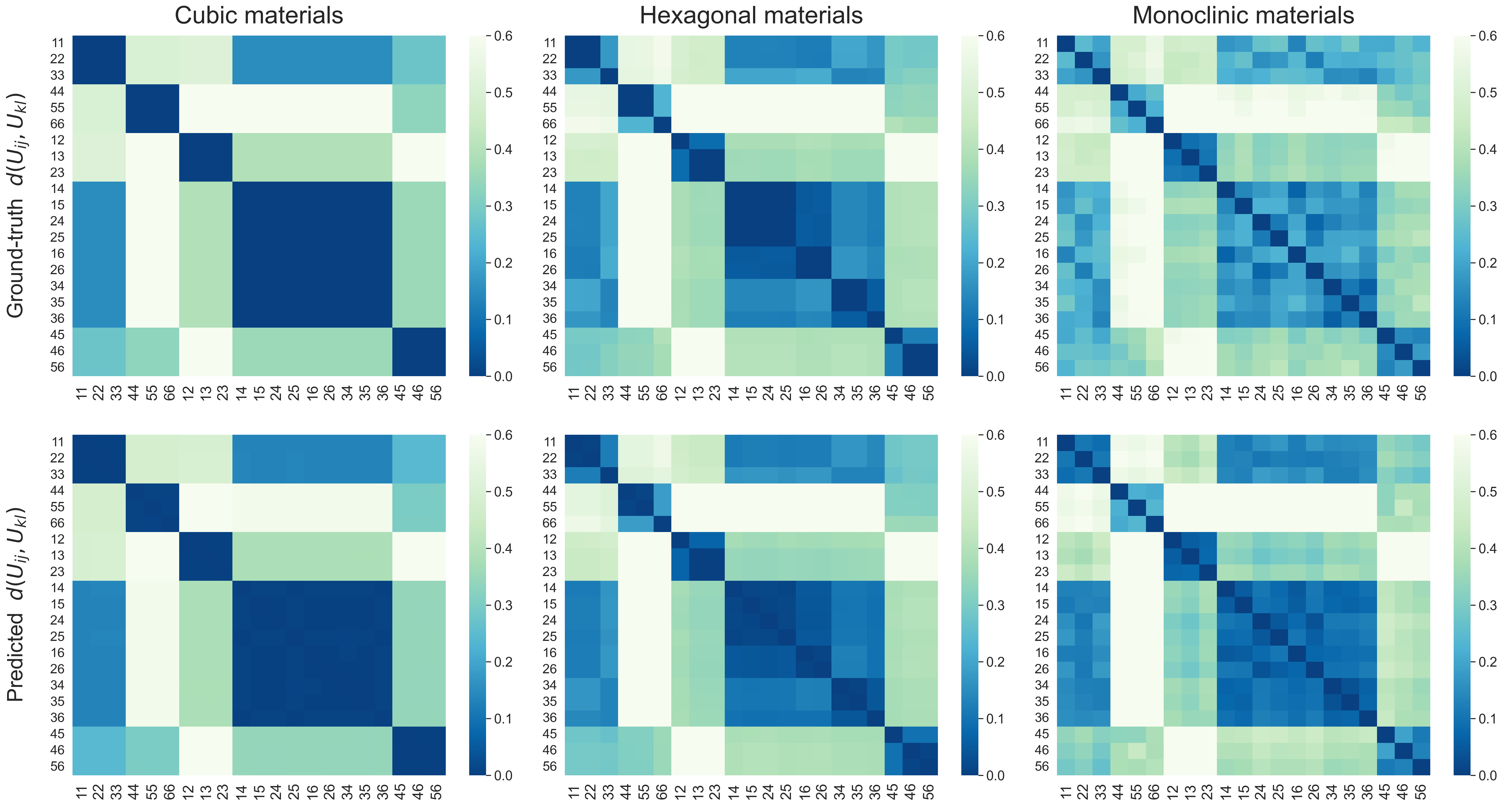}
\caption{\label{fig:distance_metric} The distance metrics $d(U_{ij},U_{kl})$ computed from the test set of the ground-truth SED (top row) and the predicted SED (bottom row). The columns organize the crystal structure, appearing from left to right is for cubic, hexagonal, and monoclinic materials respectively. The model prediction shows an excellent agreement with the ground-truth for crystals with high symmetry (cubic), and a good agreement with the ground truths for crystals with lower symmetry (hexagonal and monoclinic).}
\end{figure*}

The crystal graph of a strained crystal structure is an input feature of the SE(3)-equivariant GNNs building block, which generates {\it two} latent features. The first latent feature (${\bf f}^0_{\textrm{class},n}$ in Fig.~\ref{fig:model}) is fed into a {\it classification} neural network that predicts a vector of dimension $21$, identifying all the upper-triangular elements of the SED tensor whose values are exactly identical (by symmetry) to the component indicated by the input one-hot vector. This inherent classification informs the model to discern the degenerate structure of the SED tensor that depends on the symmetry of an input crystal graph (see Fig.~\ref{fig:distance_metric}). For example, for a cubic lattice strained in the direction $11$, the input one-hot vector $(1,0,\dots,0)^T$ together with the latent feature ${\bf f}^0_{\textrm{class},n}$ shall give a prediction of a vector that has the value close to 1 in the indices $ij = 11, 22, 33$, and the value close to 0 in the other 18 indices. Finally, the predicted {\it degeneracy class vector} (DCV) together with the second latent feature (${\bf f}^0_{\textrm{reg},n}$ in Fig.~\ref{fig:model}) will be fed into the final neural network that predicts (regresses on) the SED.


To generate an SE(3)-equivariant latent representation of an input crystal graph, as alluded to in the previous section, we first used the SE(3)-Transformer that receives the strained crystal graph input. 
The material structure from the database is first transformed according to Eq.~\eqref{eq:lattice} by a strain of a fixed magnitude, which is then converted into the strained crystal graph input. Noting that in the \textit{ab initio} calculation, the atomic coordinates in the strained lattice are optimized to be in the equilibrium positions under an applied strain, but in this work, the fractional coordinates are kept as in pre-strain condition.  The strained crystal graph input is given by 
\begin{eqnarray*}
    &\mathcal{V}' = \{(\mathbf{f}_n, \mathbf{r}^{\prime}_n) \; | \; \mathbf{f}_n \in \mathbb{R}^{M}, \; \mathbf{r}^{\prime}_n =  \bm{\upvarepsilon}_I\mathbf{L}\mathbf{r}_{f_n} \in \mathbb{R}^{3} \}   , \\
    &\mathcal{E}' = \{\Delta\mathbf{r}_{mn}^{\prime (\mathbf{T}^{\prime})} \; | \; \Delta\mathbf{r}_{mn}^{\prime (\mathbf{T}^{\prime})} = \mathbf{r}^{\prime}_m - \mathbf{r}^{\prime}_n + \mathbf{T}^{\prime}; \ \mathbf{r}^{\prime}_m, \mathbf{r}^{\prime}_n \in \mathbb{R}^{3} \} ,
\end{eqnarray*}
where the \textit{prime} notation indicates that the atomic positions and the translation vector are of the strained lattice. 

 The training data of the $\{U_{ij}\}$ is computed, via  Eq.~\eqref{eq:uij}, from $\{C_{ij}\}$ extracted from Materials Project database \cite{Jain2013} (see Appendix ~\ref{sec: dataset}).  
For molecules and materials, if the number of training data is large enough, only atomic numbers are sufficient for the node (atomic) attributes \cite{Fung2021}. Nevertheless, in this work,
the features such as atomic mass, electronegativity, polarizibility, atomic radius, atomic number, and $ij$ indices, are used as node attributes. Polarizability data is obtained from \cite{Schwerdtfeger2018}, whereas other attributes are obtained from Materials Project database. The atomic number is vectorized through an embedding layer of dimension 512, which is then concatenated with the other four node attributes (see Fig.~\ref{fig:model}). 
\textcolor{black}{Note that the elastic constants are influenced by the atomic mass since the phonon frequencies in sound waves, particularly the acoustic modes, are inversely proportional to the square root of atomic mass.}
Additionally, the resistance to changes in bond length correlates with both electronegativity and polarizability. These three atomic properties are thus incorporated into node attributes, and they significantly improve the model prediction accuracy.

Since the atomic features are scalar attributes, the input feature vector will be fed into the $l=0$ channel of the transformer network since such channel transforms like a scalar (see Appendix.~\ref{sec: SE(3)-transformers}). We'll denote each node's input feature vector as ${\bf f}^{0}_{\textrm{in},n} = {\bf f}_{n}$.
For computational feasibility, we will restrict the output feature vector of the SE(3)-transformers to consist of 4 channels ${\bf f}^{l}_{\textrm{hid},n}$ with $l\in\{0,1,2,3\}$. These latent features are fed into two different SE(3)-equivariant GNNs {\it without} attention (TFNs), outputing a node-wise classification feature vector ${\bf f}^{0}_{\textrm{class},n}$, and the node-wise regression feature vector ${\bf f}^{0}_{\textrm{reg},n}$, see Fig.~\ref{fig:model}. Note that the attention mechanism is not deployed in the last GNN layer as it yields a better prediction accuracy. 

To train the model to discern different orientations of a strained crystal graph and classify its appropriate degeneracy class vector, in each epoch, we draw a new realization of a crystal graph in a different orientation and of a new one-hot vector in the same class of vector associated with the original strained component.  Specifically, a new crystal graph input is sampled from a uniformly random orientation (uniformly random Euler angles) around the center of mass in the unit cell of the original strained crystal graph.  A new one-hot vector of the strained components is chosen such that the element 1 is drawn uniformly from one of the non-zero components of the DCV. For example, if the original strained component is $11$ for a cubic material, a new one-hot vector in each epoch is drawn from the situations where only one of the strained components 11, 22, or 33 is 1, while the other components are 0. 

Our self-supervised approach combines both the regression errors and the classification errors in the global loss function:
\begin{equation}
    \mathcal{L} = \frac{1}{N}\sum_{n=1}^N|U_{ij}^{(n)} - U_{ij}^{(n),\textrm{pred}}| + \lambda \mathcal{L}_{\textrm{class}},
\end{equation}
where the SED prediction $U_{ij}^{(n), \textrm{pred}}$ is the output from the regression network,  $\mathcal{L}_{\textrm{class}}$ is the binary cross-entropy loss function for multi-label classification, $\lambda$ specifies the relative significance between the classification and the regression errors, and  $N$ is the total number of training samples. We stop the training when the gradient of the loss is relatively small over multiple epochs. Training was performed on NVIDIA A100 GPUs.0

\subsection{Choosing a subset of strained crystal structures for training data}
By the virtue of rotational equivariance, we expect \textcolor{black}{StrainTensorNet} to efficiently predict the 21 components of $\mathbf{U}$ using training data consisting of only non-degenerate components. We have selected the number of non-degenerate components for training that depends on crystal systems, i.e. cubic, tetragonal, hexagonal, trigonal, monoclinic, and triclinic. For cubic materials, only 6 components are used for training (the number of non-degenerate components for other crystal systems can be found in Table~\ref{tab:crystal_system}). Recall that, as discussed earlier, the cubic lattice has 5 distinct $U_{ij}$ as shown in Eq.~\eqref{eq:u_cubic}; however, the strain tensor transform the cubic structures into 6 distinct structures (according to the symmetry in Laue class $m\bar{3}m$) as follows
\begin{equation*}
    \begin{bmatrix}
    \mathcal{T}_1&\mathcal{T}_2&\mathcal{T}_2&\mathcal{O}_2&\mathcal{M}_2&\mathcal{M}_2 \\
    &\mathcal{T}_1&\mathcal{T}_2&\mathcal{M}_2&\mathcal{O}_2&\mathcal{M}_2 \\
    &&\mathcal{T}_1&\mathcal{M}_2&\mathcal{M}_2&\mathcal{O}_2 \\
    &&&\mathcal{O}_1&\mathcal{M}_1&\mathcal{M}_1 \\
    &&&&\mathcal{O}_1&\mathcal{M}_1 \\
    &&&&&\mathcal{O}_1
    \end{bmatrix},
\end{equation*}
where each element in the matrix stands for a crystal system of $\mathbf{L}'(\epsilon_i,\epsilon_j)$, $\mathcal{T}$, $\mathcal{O}$, and $\mathcal{M}$ are tetragonal, orthorhombic, and monoclinic lattices, respectively, and a different number labeling the subscript indicates a different structure. Despite the fact that $U_{14}$ ($C_{14}$) is equal to $U_{15}$ ($C_{15}$) by the cubic symmetry, $\mathbf{L}'(\epsilon_1,\epsilon_4)$ and $\mathbf{L}'(\epsilon_1,\epsilon_5)$ possess different lattice symmetries which are orthorhombic and monoclinic lattices, respectively. This is because $\epsilon_1$ strains the structure in the $x$ direction, while $\epsilon_4$ and $\epsilon_5$ strain the structure in both the $y$ and the $z$ directions, and both the $x$ and the $y$ directions, respectively. Since $\mathbf{L}'(\epsilon_1,\epsilon_4)$ and $\mathbf{L}'(\epsilon_1,\epsilon_5)$ are not equivalent up to a rotation, the SE(3) kernel regards the two inputs as different inputs. By exploiting SE(3) kernel to identify inputs that are equivalent up to a rotation, for cubic materials, it suffices to use the training data consisting of the distinct input structures, strained only in 6 directions, i.e., 11, 12, 14, 15, 44, and 45. 

\subsection{The distance between SED components}
To evaluate our model capability to predict the symmetry-dependent degeneracy pattern of the SED tensor, we use the Canberra distance between two components of the SED tensor as a metric:
\begin{equation}\label{eq:distance}
    d(U_{ij}, U_{kl}) \equiv \frac{1}{N}\sum_{n=1}^N\frac{|U_{ij}^{(n)} - U_{kl}^{(n)}|}{|U_{ij}^{(n)}| + |U_{kl}^{(n)}|},
\end{equation}
where $N$ represents the number of samples, and $ij$ and $kl$ are Voigt indices. With this metric, the distance between $U_{ij}$ and $U_{kl}$ is zero if they belong to the same degeneracy class (their values are identical.) The top row of Fig.\ref{fig:distance_metric} shows the ground-truth distance pattern for cubic, hexagonal, and monoclinic materials. Dark blue and bright green colors indicate the $d(U_{ij},U_{kl})$ values that are closest to zero and 0.6, respectively. It is important to note that the degeneracy pattern of $U_{ij}$ depends on the crystal symmetry, which results in a unique Canberra distance pattern for each crystal system. For instance, in cubic materials, $U_{11}$, $U_{22}$, and $U_{33}$ are identical (in the same degeneracy class), so the distance pattern of cubic materials displays a dark blue box corresponding to indices 11, 22, and 33.
The distance patterns in Figs.~\ref{fig:distance_metric} and~\ref{fig:dm_other} are averaged over the samples with the same crystal system in the dataset. Hence, the distance pattern of low-symmetry crystals, such as monoclinic and triclinic, will strongly depend on the dataset.

\section{Discussion and Results}\label{sec: results}

\subsection{Rationale for the \textcolor{black}{StrainTensorNet} model architecture}
The underlying principle for using an equivariant network as a core component is to construct a compact representation of the SED tensor, such that the crystal structures strained by different sets of $\epsilon_i$ and $\epsilon_j$ that are equivalent up to rotation (and translation) possess the same SED. Such symmetry-dependent SED tensor of specific crystal structures is demonstrated in the ground-truth distance patterns $d(U_{ij},U_{kl})$ of Fig.~\ref{fig:distance_metric} (top), where the \textcolor{black}{StrainTensorNet} can well approximate such symmetry-dependent patterns, see Fig.~\ref{fig:distance_metric} (bottom).

Fig.~\ref{fig:additional_models} shows our earlier model development with the goal to compactly represent the SED tensor, beginning from the simplest but less expressive to \textcolor{black}{StrainTensorNet} which can compactly represent the SED tensor rather accurately. All these trial GNN-based models were trained on crystal structures with fewer than 500 edges. The simplest model (Fig.~\ref{fig:additional_models}(a)) takes as an input only the crystal graph with node attributes, i.e., atomic mass, atomic radius, electronegativity, polarizability, and embedded atomic number. This minimal equivariant model does not yield a sufficiently accurate representation of $U_{ij}$; we found that the model is more biased towards predicting the SED tensor whose distance matrix pattern is more akin to that of the cubic crystals. This could be because the model is not sufficiently expressive, and thus attempts to fit the majority of the training dataset comprising more higher-symmetry structures (see Table~\ref{tab:crystal_system_data} for the crystal system statistics of our curated dataset).

To improve the minimal model's expressiveness, we introduce the degeneracy class vector (DCV) as an additional input, to inform the model about the symmetry-dependent class of the SED tensor. The DCV is first embedded by fully-connected layers and then concatenated with global max-pooled features or node attributes, as shown in Fig.~\ref{fig:additional_models}(b) and (c), respectively. These symmetry-informed models yield lower MAE and RMSE of the SED tensor compared to that of the minimal model, while the model in Fig.~\ref{fig:additional_models}(b) performs slightly better than the model in Fig.~\ref{fig:additional_models}(c)  (see Table~\ref{tab:models_bcd}). Moreover, the predicted distance patterns are less biased towards those of the cubic lattice. Specifically, for crystal structures with lower symmetry such as hexagonal, tetragonal, trigonal, and orthorhombic, the predicted distance patterns are relatively similar to their ground-truth distance patterns. As expected, informing the model about the DCV of the input helps force the distance between $U_{ij}$ and $U_{kl}$ belonging to the same degeneracy class to be closer to zero.

\begin{figure*}[h]
\includegraphics[width=0.78\textwidth]{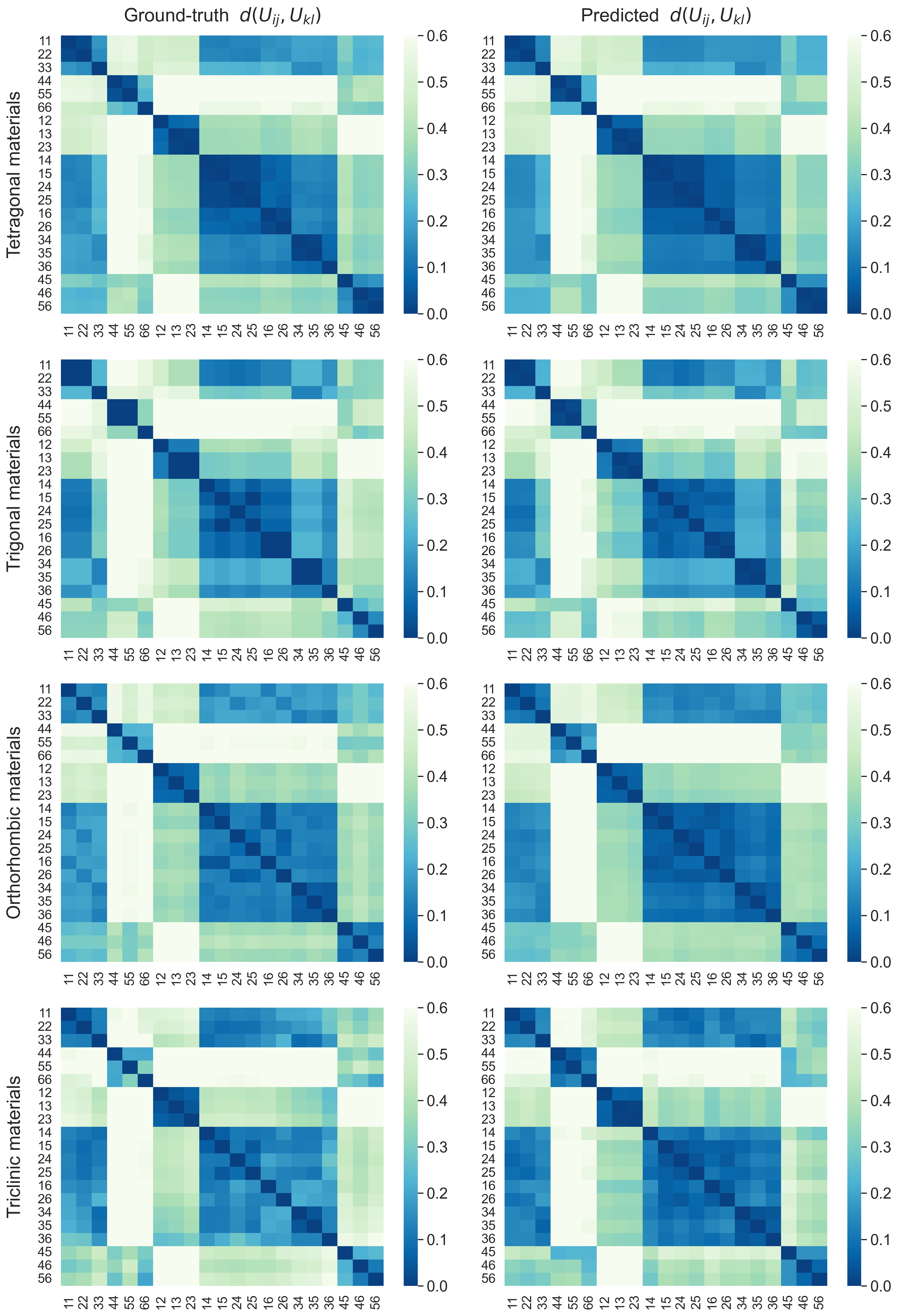}
\caption{\label{fig:dm_other} The distance metric (computed from the test set) of tetragonal, trigonal, orthorhombic, and triclinic materials (top to bottom respectively) for (left) the ground-truth and (right) \textcolor{black}{StrainTensorNet}'s prediction.}
\end{figure*}

While these models with a DCV included as an input give good predictions of the SED tensor, they require the knowledge of the prestrained crystal symmetry {\it a priori} (to properly assign the value of the DCV). To overcome this limitation, we use a self-supervised method schematically shown Fig.~\ref{fig:model} to modify the model to  predict, rather than to require, the DCV. The classification neural network that predicts the DCV takes as an input a one-hot vector of the strained $ij$ components, representing the $\epsilon_i$ and $\epsilon_j$ components in ~\eqref{eq:strain_matrix}. This self-supervised technique enables the \textcolor{black}{StrainTensorNet} to predict $U_{ij}$ without any prior knowledge of the prestrained crystal symmetry. Importantly, \textcolor{black}{StrainTensorNet} can express the degeneracy class of the SED tensor relatively well, see Figs.~\ref{fig:distance_metric} and~\ref{fig:dm_other}.

\subsection{\textcolor{black}{StrainTensorNet}'s prediction results}\label{sec:pred_result}

\begin{figure*}[h]
\includegraphics[width=1.0\textwidth]{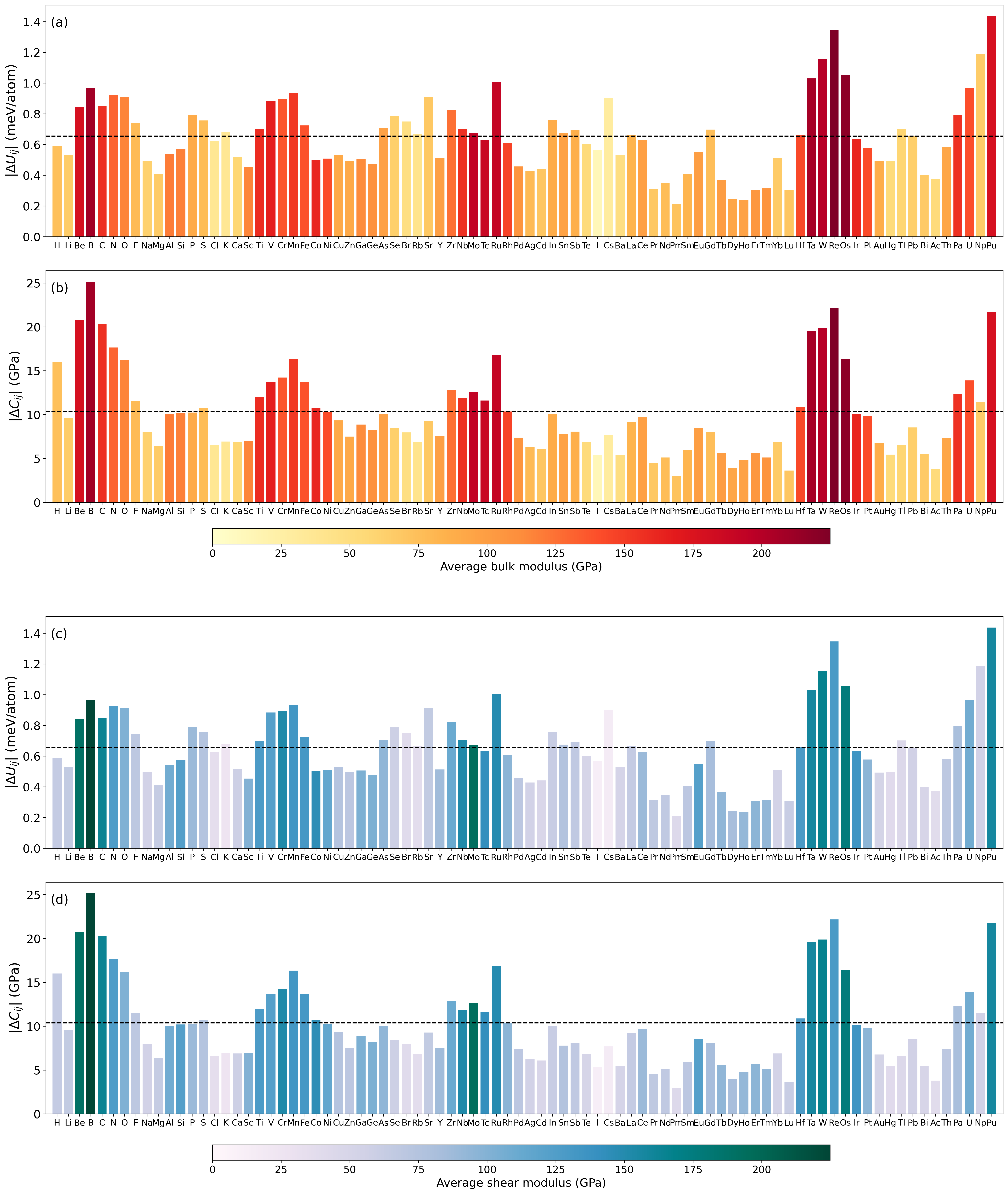}
\caption{\label{fig:errors_u_and_c} 
The bar charts display the MAE of the predicted $U_{ij}$ and $C_{ij}$ components, averaged over 21 $ij$ components. The charts are categorized based on the elements present in the compounds in the test set and are colored according to the average bulk modulus for (a) and (c), and average shear modulus for (b) and (d), as indicated in the legend below. The dashed lines represent the MAE of each property, averaged over the data in the test set. It can be seen that most elements, whose compounds have high bulk and shear moduli, exhibit errors of $U_{ij}$ and $C_{ij}$ that exceed the dataset's MAE.
}
\end{figure*}

Table~\ref{tab:table1} summarizes \textcolor{black}{StrainTensorNet}'s prediction errors on the test set. To the best of our knowledge, \textcolor{black}{StrainTensorNet} provides the first data-driven prediction of $U_{ij}$ with a reasonable accuracy. Using these predicted SED tensors together with Eq.~\eqref{eq:cij}, elastic tensors of materials can also be calculated. Our model's prediction accuracy significantly depends on the elemental composition of the compounds. Fig.~\ref{fig:errors_u_and_c} depicts the MAE of $U_{ij}$ and $C_{ij}$, averaged over 21 $ij$ components and categorized by elements. It is interesting to note that the compounds containing period 2 elements (Be, B, C, N, O, and F), some period 6 transition metals (Ta, W, Re, and Os), and some actinides (Th, Pa, and Pu), which have high bulk and shear moduli on average, exhibit a higher MAE for $U_{ij}$ compared to the overall MAE of the entire test set. As a result, their MAE for $C_{ij}$ are also significantly higher than the overall MAE of the test set. In contrast, the compounds containing lanthanides, which exhibit medium bulk and shear moduli on average, except for Gd, demonstrate a considerably lower MAE than the overall MAE of the test set, for both $U_{ij}$ and $C_{ij}$.

\begin{figure*}[t]
\includegraphics[width=1.\textwidth]{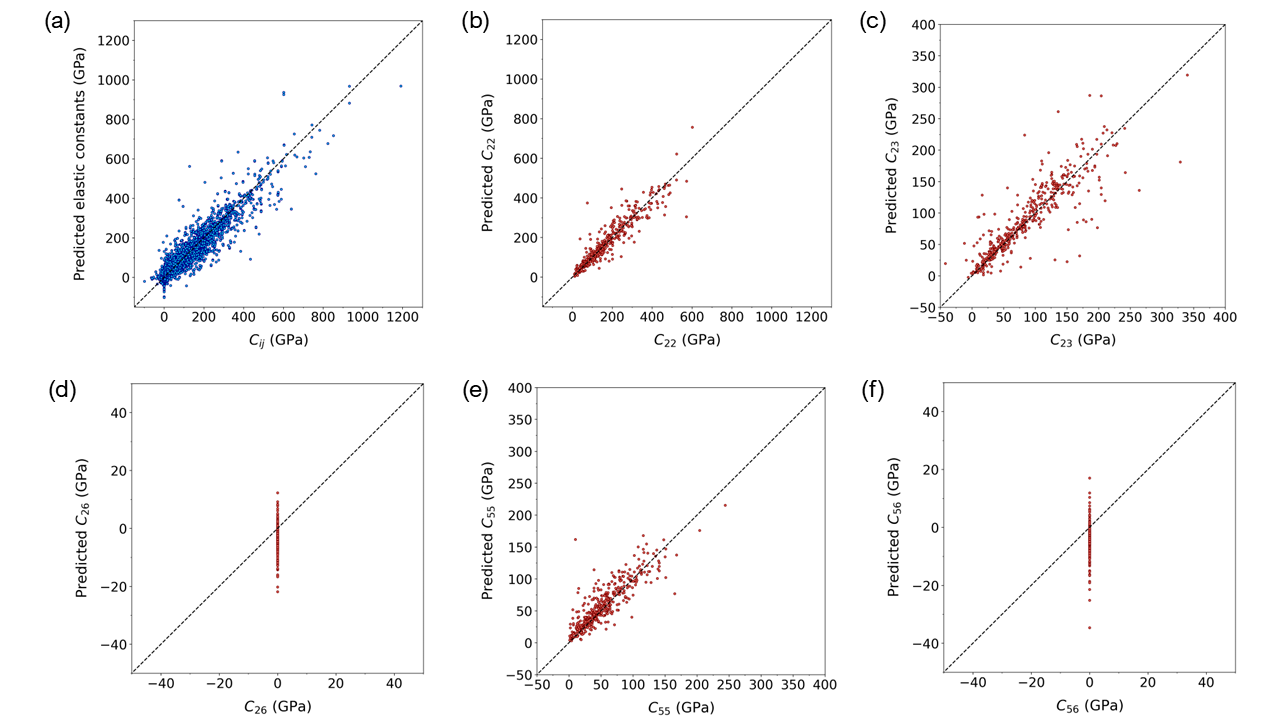}
\caption{\label{fig:scatters} The predicted values (vertical axes) versus the ground-truth values (horizontal axes) of the elastic constants.  (a) All the components of the elastic constants of the whole test set of all crystal systems. The results for only {\it cubic materials} are shown in (b) $C_{22}$, (c) $C_{23}$, (d) $C_{26}$, (e) $C_{55}$, and (f) $C_{56}$, where we note that the 22, 23, 26, 55, and 56 indices {\it did not} appear in the input of the training set for cubic materials. }
\end{figure*}

To further investigate the prediction results of the elastic constants, we plot the comparisons between the prediction (obtained from the predicted SED tensors together with Eq.~\eqref{eq:cij}) and the ground truth of the elastic constants in Fig.~\ref{fig:scatters}. Fig.~\ref{fig:scatters} (a) shows the prediction results of every $C_{ij}$ components of all crystal structures,  whereas Figs.~\ref{fig:scatters}(b)-(f) show the prediction results of $C_{22}$, $C_{23}$, $C_{26}$, $C_{55}$, and $C_{56}$ of cubic materials. For cubic materials, the model is trained only on a non-degenerate subset of 21 $U_{ij}$ components, specifically components 11, 12, 14, 15, 44, and 45. Notably, the model is able to predict other unseen components reasonably well. However, the prediction of the elastic constant components that are exactly zero is challenging. For example, the ground-truth values of $C_{26}$ and $C_{56}$ of cubic materials are exactly 0 GPa, while the means and standard deviations of these components from our model predictions are 0.69 and 4.45 GPa, and -0.43 and 4.85 GPa respectively (see Fig.~\ref{fig:jointplots} for the histograms of the prediction results). Although the standard deviations are relatively large, the means are less than 1 GPa. It is worth noting that the first-principles calculations also make some errors in these values, but these are usually not computed since they are known to be exactly 0 GPa.

To see how well \textcolor{black}{StrainTensorNet} discerns the geometric relationship between different strained crystal graph inputs, we plot the Canberra distance metric of the predicted and ground-truth SED tensors for comparison in Fig.~\ref{fig:distance_metric}. The predicted distance metric of cubic materials, as shown in the left column of Fig.~\ref{fig:distance_metric}, closely resembles the ground-truth distance metric, although there are minor discrepancies in the values that belong to the same degeneracy class. For instance, the distances between the predicted $U_{11}$, $U_{22}$, and $U_{33}$ are $4.8-6.2$ $\mu$eV/atom, resulting in the percentage errors of $1.83-2.43\%$  (computed from $\frac{1}{N}\sum_{n=1}^N|U_{ij}^{(n)} - U_{kl}^{(n)}|/U_{avg}^{(n)}\times100\%$ where $U_{avg}^{(n)} = (|U_{ij}^{(n)}| + |U_{kl}^{(n)}|)/2$.) For hexagonal materials, $U_{11}$ are identical to $U_{22}$, but not to $U_{33}$. The predicted distance metric of hexagonal materials accordingly forms a dark blue box in the 11 and 22 components. The percentage error between the predicted $U_{11}$ and $U_{22}$ is 2.22\%. For monoclinic materials, the values of each $U_{ij}$ are not necessarily identical to one another, and the distance metric is averaged over monoclinic materials data. The resulting distance pattern may differ significantly depending on the dataset. Notably, these agreements between the predicted and the ground-truth distance metric in various crystal structures reveal that \textcolor{black}{StrainTensorNet} gives an excellent prediction of the degenerate structure of SED tensors of strained crystalline materials.

\clearpage

Lastly, the model can also predict other elastic properties, including Voigt's bulk modulus ($B_V$), Voigt's shear modulus ($G_V$), Young's modulus ($Y$), and Poisson's ratio ($\nu$), using the predicted $C_{ij}$ together with Eqs.~\eqref{eq:bulk}$-$\eqref{eq:possion}. Table~\ref{tab:table1} summarizes the MAE and RMSE of our predicted elastic properties compared to previous works. The MAE and RMSE of $B_V$ are 12.59 and 20.83 GPa, respectively, which are comparable to those of Mazhnik et al. \cite{Mazhnik2020}. The MAE and RMSE of $G_V$ are 10.49 and 18.95 GPa, respectively, with the RMSE of $G_V$ being comparable to that reported by Zhao et al. \cite{Zhao2020}. Our work also yields smaller errors for $\nu$ than Mazhnik et al. \cite{Mazhnik2020}
Note that our dataset (see Sec. \ref{sec: dataset}) differs from the dataset in the relevant work of \cite{Mazhnik2020, Zhao2020}. Fig.~\ref{fig:errors_probs} presents the MAE of $B_V$, $G_V$, $Y$ and $\nu$, categorized by elements. The model produces high MAE for compounds containing  period 2 elements, mainly due to their larger MAE in $U_{ij}$ and $C_{ij}$ (see Fig.~\ref{fig:errors_u_and_c}). On the other hand, the model can produce smaller MAE for lanthanide compounds, as their MAE in $U_{ij}$ and $C_{ij}$ (see Fig.~\ref{fig:errors_u_and_c}) are smaller.

\subsection{Interpretability of \textcolor{black}{StrainTensorNet}'s latent features}\label{sec: ML-latent}

\begin{figure*}[t]
\includegraphics[width=1.0\textwidth]{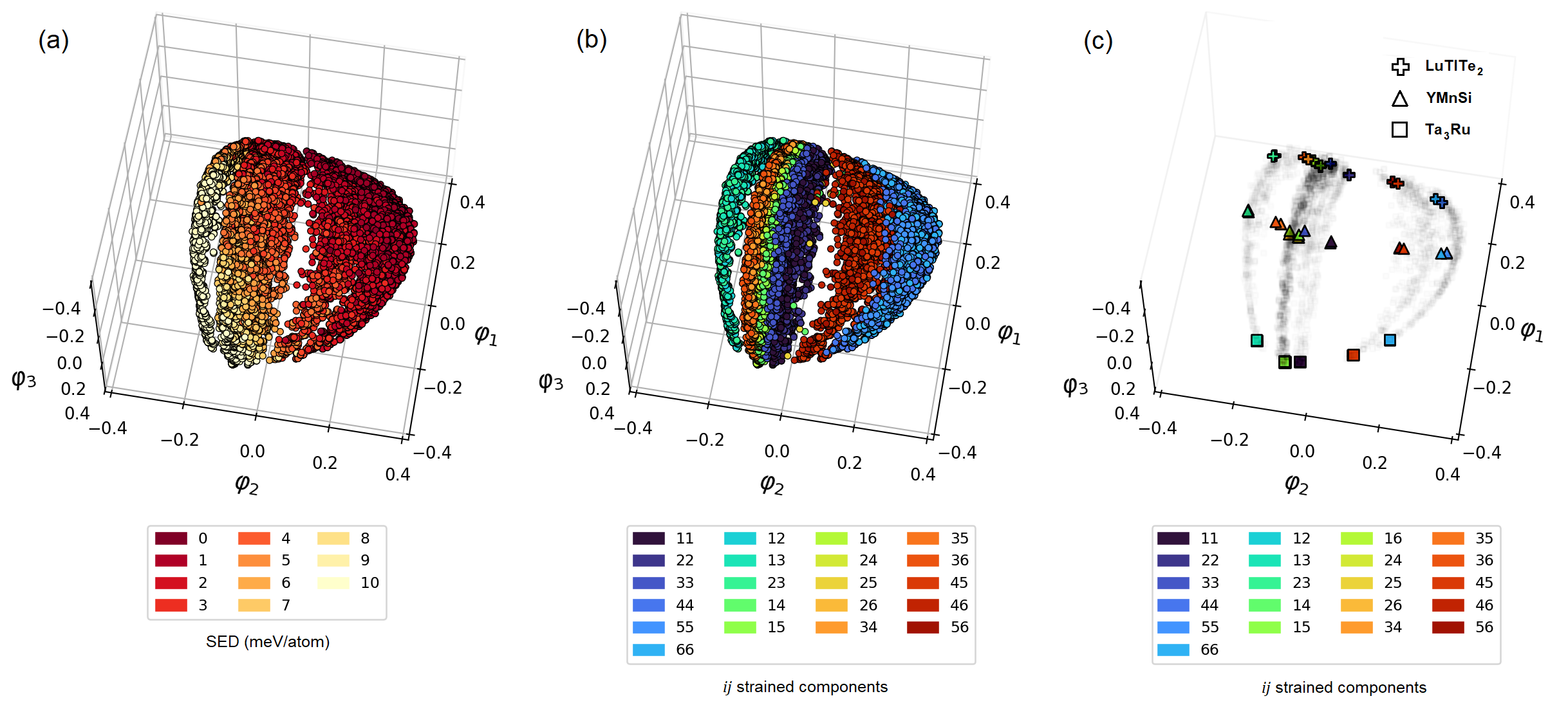}
\caption{\label{fig:diffmap} 
Three-dimensional diffusion map representation of the concatenated latent features that are fed into the SED tensor prediction layer (the latent feature \textbf{B}  in Fig.~\ref{fig:model}). (a) is colored by the SED scale, and (b) is colored by the strained component $ij$. (c) highlights the latent features of 3 example materials colored by the strained component $ij$ as in (b); the features of other materials are shown in transparent grey. The 3 materials are trigonal LuTlTe$_2$ phase ($\RPlus$), tetragonal YMnSi phase ($\triangle$), and cubic Ta$_3$Ru phase ($\square$), which have low, medium, and high elastic constants, respectively. These are all the latent features in the test dataset. 
More elaborated discussion can be found in Sec.~\ref{sec: ML-latent}.
} 
\end{figure*}

\begin{figure*}
\includegraphics[width=1.\textwidth]{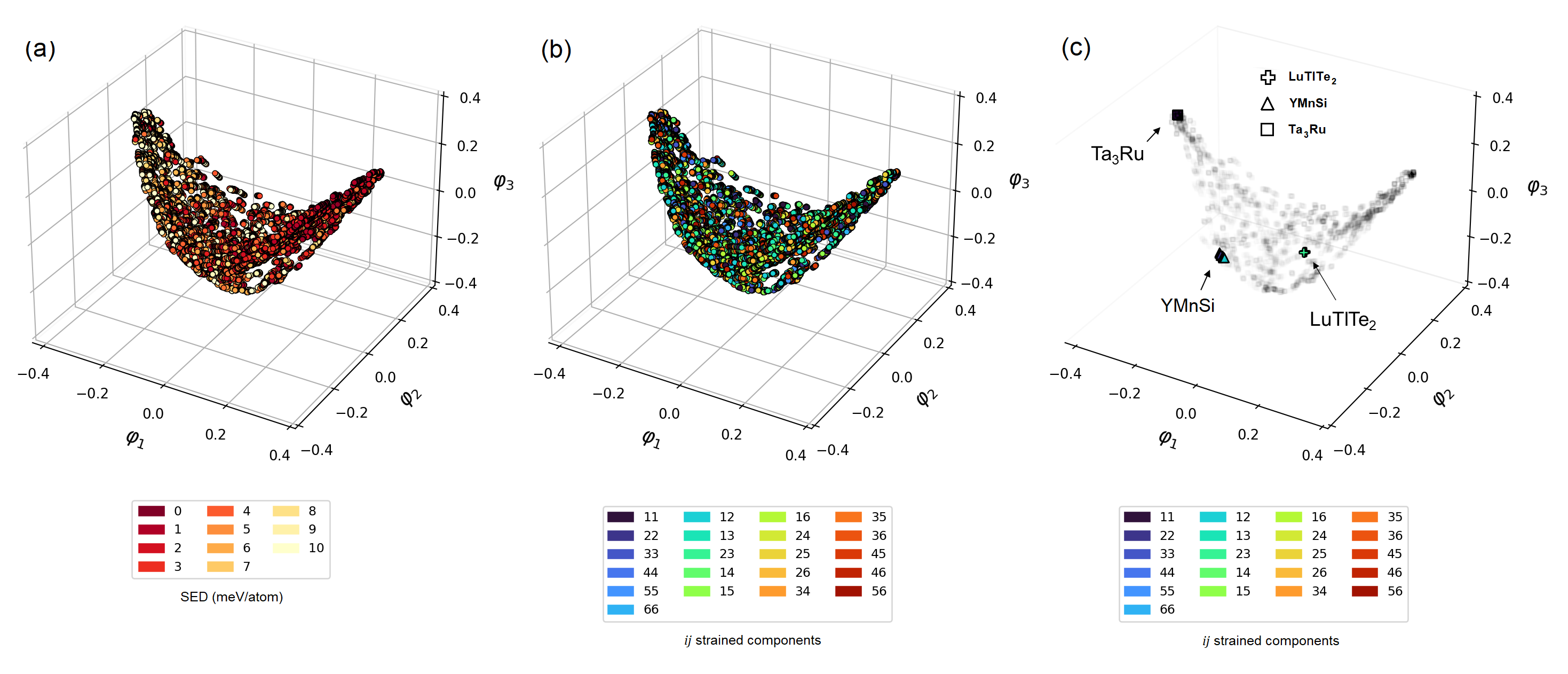}
\caption{\label{fig:reg_pool} 
Three-dimensional diffusion map representation of the global max-pooled latent features of the graph neural networks (i.e., the latent feature \textbf{A} in Fig.~\ref{fig:model}). (a) is colored by the SED scale, and (b) is colored by the strained component $ij$. (c) highlights the latent features of 3 example materials colored by the strained component $ij$ as in (b); the features of other materials are shown in transparent grey. The 3 materials are trigonal LuTlTe$_2$ phase ($\RPlus$), tetragonal YMnSi phase ($\triangle$), and cubic Ta$_3$Ru phase ($\square$), which have low, medium, and high elastic constants, respectively. See the discussion and the interpretation of these plots in Sec.~\ref{sec: ML-latent}.
 }
\end{figure*}

To investigate how latent features facilitate successful prediction of the SED tensor, we employed the diffusion map to faithfully visualize the three-dimensional data representation of the high-dimensional latent features of the entire test set \cite{lafon2004diffusion, coifman2006diffusion}. Figs.~\ref{fig:diffmap} and~\ref{fig:reg_pool} reveal the dimensionality reduction of the latent features in the diffusion coordinates $(\varphi_1, \varphi_2, \varphi_3)$. Two different latent features were considered: the global max-pooling layer from the GNN representation of the input crystal graph (denoted as \textbf{A} in Fig.~\ref{fig:model}), and the concatenation between \textbf{A} and the embedded  DCV (denoted as \textbf{B} in Fig.~\ref{fig:model}). These low-dimensional representations in Figs.~\ref{fig:diffmap} and~\ref{fig:reg_pool} are colored by the energy scale (left column) and by the strained component ${ij}$ (middle and right columns).

What does the max-pooled feature of the crystal graph input \textbf{A} represent? Fig.~\ref{fig:reg_pool}(c) shows that the crystal graph latent representations of the {\it same material} that are strained in different $ij$ components are {\it almost identical}, in fact sharing greater resemblance than the representations of different materials that belong to the same symmetry class. This is consistent with the result of Bronstein et al. which shows that the latent representation in SE(3)-equivariant GNNs of the original input is very similar to those of the mildly spatially distorted inputs, and, interestingly, is less similar to those of inputs spatially translated further away \cite{Bronstein2021}. Hence, the latent feature \textbf{A} seems to uniquely encode the information of the input material, rather than the symmetry of the input. Thus \textbf{A} alone does not suffice to meaningfully represent the symmetry-dependent SED tensor (Fig.~\ref{fig:reg_pool}(a) and (b)).

On the other hand, with the embedded DCV in the latent feature \textbf{B}, the network can differentiate different strained components within the same material class.  When strained in 21 different components, the latent features \textbf{A} of three example materials (trigonal LuTlTe$_2$ phase, tetragonal YMnSi phase, and cubic Ta$_3$Ru phase) that were not differentiable within the same material class (Fig.~\ref{fig:reg_pool}(c)) are now clearly segregated in the $\varphi_2$ variable through the help of the embedded DCV, see Fig.~\ref{fig:diffmap}(c).  In fact, the latent feature \textbf{B} which combines both the input crystal graph representation and the embedded DCV offers an interpretable latent representation of the SED (Fig.~\ref{fig:diffmap}), which we now discuss. 

The latent feature \textbf{B} is organized such that the $(\varphi_1,\varphi_2)$ coordinate encodes the strain energy density that varies from a smaller value in the $(+,+)$ quadrant to a larger value in the $(-,-)$ quadrant (see Fig.~\ref{fig:diffmap}(a)). Since $U_{ij}$ stored by the tensile strain  ($i,j\in\{1,2,3\}$) is typically higher than $U_{ij}$ stored by the shear strain  ($i,j\in\{4,5,6\}$), Figs.~\ref{fig:diffmap}(a) and (b) consistently reveal that the latent features of the materials strained by the $11$, $22$, or $33$ component have a negative $\varphi_2$ value, whereas those of the materials strained by the $44$, $55$, or $66$ components have a positive $\varphi_2$ value. In addition, since $U_{ij}$ for $i\neq j$ is computed from the sum of $C_{ii}$, $C_{jj}$, and $C_{ij}$, it is typically larger than $U_{ii}$ and $U_{jj}$. Figs.~\ref{fig:diffmap}(a) and (b) also consistently reveal that the latent features of the materials strained by 44, 45, 11, and 12 components are respectively organized in the $\varphi_2$ coordinate from a more positive to a more negative value. Additionally, Figs ~\ref{fig:diffmap}(a) and (c) show that materials with higher (lower) average SED will be represented by a larger negative (larger positive) $\varphi_1$ variable. In summary, the max-pooled feature from the graph neural networks \textbf{A} encodes the material information, and, together with the information of the DCV, the concatenated latent feature \textbf{B} effectively encapsulates both the material information and its strained component-dependent SED.

\clearpage

\begin{table*}
\caption{\label{tab:table1} Statistics of the dataset and the model prediction}
\begin{ruledtabular}
\begin{tabular}{cccccccc}
\multirow{2}{*}{Properties}&\multicolumn{3}{c}{This work}&\multicolumn{3}{c}{Mazhnik et al. \cite{Mazhnik2020}}&Zhao et al. \cite{Zhao2020}\\
& Average & MAE & RMSE & Average & MAE & RMSE & RMSE \\
\hline
$U_{ij}$ (meV/atom) & 2.652 & $\mathbf{0.655}$ & $\mathbf{1.288}$\\
$C_{ij}$ (GPa) & 42.92 & $\mathbf{10.37}$ & $\mathbf{17.69}$\\
$B_V$ (GPa) & 107.07 & 11.48 & 19.73 & 111.83 & 11.11 & 19.54 & 16.530\\
$G_V$ (GPa) & 50.98 & 9.61 & 17.10 & 54.81 & 8.24 & 11.43 & 15.780\\
$Y$ & 129.62 & 22.29 & 38.63 & 138.95 & 19.15 & 26.23\\
$\nu$ & 0.401 & $\mathbf{0.037}$ & 0.133 & 0.286 & 0.041 & 0.105
\end{tabular}
\end{ruledtabular}
\end{table*}

\section{Conclusions}\label{sec: conclusion}
We have demonstrated a novel data-driven framework for predicting the elastic properties of crystal structures using SE(3)-equivariant graph neural networks (GNNs). 
By leveraging SE(3)-equivariant GNNs as building blocks, our self-supervised deep learning model accurately predicts Voigt's bulk modulus, Voigt's shear modulus, Young's modulus, and Poisson's ratio that are comparable to other non-equivariant deep learning studies \cite{Mazhnik2020, Zhao2020}. 

A key contribution is the prediction of the strain energy density (SED) and its associated elastic constants, which are tensorial quantities that depend on a material's crystallographic group. The similarity between the distance metrics of the SED components between the ground truths and the predictions demonstrates our model's capability to identify different symmetry groups of strained crystal structures. Requiring only a strained crystal graph and the strained component as the input, our approach offers an efficient alternative to the standard {\it ab initio} method for designing new materials with tailored elastic properties. 

The interpretability of the model is also a notable feature. The learned latent representations taking into account the degeneracy class vector are organized in a physically meaningful structure for predicting the SED tensors. This interpretability aspect enhances the transparency of model prediction, enabling the justification of whether the prediction is physically relevant.

The combination of interpretability and the consideration of crystallographic groups sets our model apart from recent data-driven methods for predicting elastic properties of materials. We hope this work is a stepping stone toward an efficient data-driven approach for materials discovery and design,  and opens up avenues for approaching more challenging tensor prediction tasks, e.g., predicting dielectric and piezoelectric tensors, which are second-rank and third-rank tensors, respectively.

\begin{acknowledgments}
This research is supported by the Program Management Unit for Human Resources and Institutional Development, Research and Innovation (Grant No. B05F650024) and by Thailand Science research and Innovation Fund Chulalongkorn University (CU\_FRB65\_ind (5)\_110\_23\_40). The authors acknowledge high performance computing resources including NVIDIA A100 GPUs from Chula Intelligent and Complex Systems Lab, Faculty of Science, and from the Center for AI in Medicine (CU-AIM), Faculty of Medicine, Chulalongkorn University, Thailand.  The authors also acknowledge the National Science and Technology Development Agency, National e-Science Infrastructure Consortium, Chulalongkorn University and the Chulalongkorn Academic Advancement into Its 2nd Century Project (Thailand) for providing computing infrastructure.
\end{acknowledgments}

\begin{appendix}
\renewcommand\thefigure{\thesection\arabic{figure}}   

\section{Ingredients of SE(3)-Transformers}\label{sec: SE(3)-transformers}
\setcounter{figure}{0}
Here we summarize the computational building block of SE(3)-equivariant GNNs with a built-in attention mechanism (SE(3)-Transformers). More detailed discussion of this deep learning architecture can be found in the original proposal \cite{Fuchs2020}. 

In an SE(3)-equivariant GNN {\it without} attention, also known as Tensor Field Network (TFN) \cite{Thomas2018}, given a multi-channel input feature of a node $j^{th}$, $\mathbf{f}_{\text{in},j}^k$, of a crystal graph, the GNN will update each crystal graph input feature by the following message-passing convolution
\begin{equation} \label{eq:message_passing}
    \mathbf{f}_{\text{out},i}^l=\mathbf{W}^{ll}\mathbf{f}_{\text{in},i}^l+\sum_{k\geq0}{\sum_{j\neq i}^{N}{\mathbf{W}^{lk}\left(\mathbf{r}_j-\mathbf{r}_i\right)\mathbf{f}_{\text{in},j}^k}\ },
\end{equation}
where $i$ and $j$ are respectively the indices of a target node and its neighbouring nodes,  $l$ and $k$ specify type-$l$ and type-$k$ feature vectors which respectively transform under SO(3) rotation according to the $(2l + 1) \times (2l+1)$ and $(2k + 1) \times (2k+1)$ Wigner-D matrices (type-0 vector is invariant under SO(3) rotation), $\mathbf{W}^{ll}$ is  the diagonal matrix whose diagonal is identically parametrized by a trainable parameter $w^{ll}$ describing a self-interaction message passing, and $\mathbf{W}^{lk}$ is a convolution kernel that propagates the channel-$k$ feature from neighboring nodes to the channel-$l$ feature of the node of interest such that 
\begin{equation} \label{eq:filter}
    \mathbf{W}^{lk}(\mathbf{r}_j-\mathbf{r}_i) =
    \sum_{J=|k-l|}^{k+l}\varphi_J^{lk}\left(||\mathbf{r}_j - \mathbf{r}_i||\right)\mathbf{W}_J^{lk}(\mathbf{r}_j - \mathbf{r}_i),
\end{equation}
with
\begin{equation} \label{eq:kernel}
    \mathbf{W}_J^{lk}(\mathbf{r}_j-\mathbf{r}_i)=\sum_{m=-J}^{J}Y_{Jm}\left(\frac{\mathbf{r}_j-\mathbf{r}_i}{||\mathbf{r}_j - \mathbf{r}_i||}\right)\mathbf{Q}_{Jm}^{lk},
\end{equation}
where $\varphi_J^{lk}\left(||\mathbf{r}||\right)$ is a learnable radial function, $Y_{Jm}$ is the $m^{\text{th}}$ component ($m$ magnetic quantum number in quantum mechanics)  of the spherical harmonic function $Y_J: \mathbb{R}^3 \rightarrow \mathbb{R}^{2 J+1}$,  $\mathbf{Q}_{Jm}^{lk}$ is a Clebsch-Gordan matrix with dimension $(2l+1) \times (2k+1)$. The SO(3)-equivariant basis kernel $\mathbf{W}^{lk}_J$ imposes a fixed functional constraint on the convolution kernel along the angular direction, whereas the radial direction is still trainable via $\varphi_J^{lk}$  

\begin{figure}[h]
\includegraphics[width=.5\textwidth]{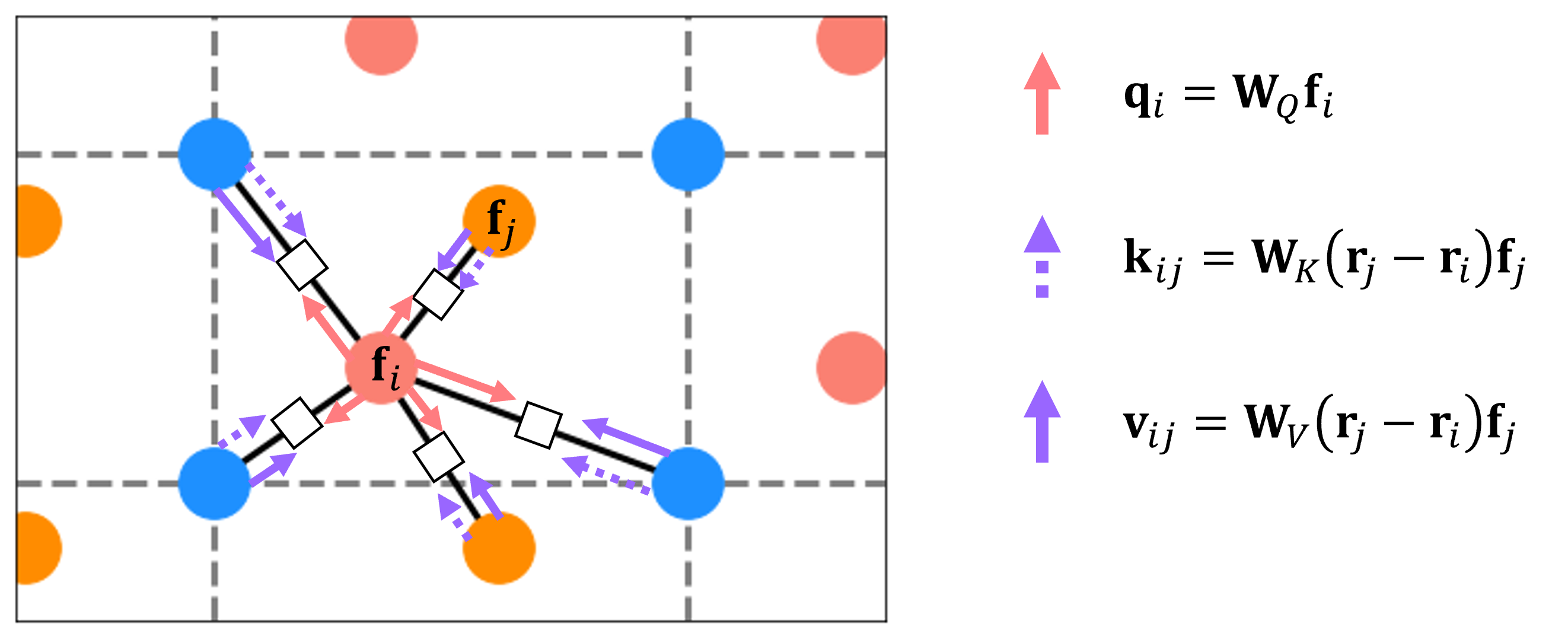}
\caption{\label{fig:se3t} A schematic of a GNN message passing rule that aggregates features from neighboring nodes of a crystal graph with an attention mechanism. A white square on each edge depicts an attention head.  A target node (pink circle) sends a query vector of the node $i$, ($\mathbf{q}_i$, pink arrow), to match with key vectors ($\mathbf{k}_{ij}$, dotted purple arrow) and the value message ($\mathbf{v}_{ij}$, purple arrow) from neighboring nodes  (blue and orange circles). The angular momentum channels are not illustrated in the visualization.}\label{fig: attention}
\end{figure}

The SE(3)-Transformer introduces the attention $\alpha_{ij}$ to each edge of a TFN, so that the message passing rule of Eq. \eqref{eq:message_passing} is weighted by the attentions. Without isolating out the self-interaction term, for brevity, this weighted message passing rule can be compactly expressed as \begin{equation}
    \mathbf{f}_{\text{out},i}^l =\sum_j\alpha_{ij}\mathbf{v}_{ij}^{l},
\end{equation}
where the channel-$l$ {\it value message}, $\mathbf{v}_{ij}^{l}$, is the same as that of the above TFN, defined as
\begin{equation}
    \mathbf{v}_{ij}^{l} = \sum_{k \ge 0}\mathbf{W}_V^{lk}(\mathbf{r}_j - \mathbf{r}_i)\mathbf{f}_{\text{in},j}^k,
\end{equation}
and the attention on the edge $ij$ (which is invariant under SO(3) \cite{Fuchs2020}) is 
\begin{equation}
    \alpha_{ij} = \frac{\mathrm{exp}(\mathbf{q}_i^{\top}\mathbf{k}_{ij})}{\sum_{j'}\mathrm{exp}(\mathbf{q}_i^{\top}\mathbf{k}_{ij'})},
\end{equation}
with a {\it query} vector on a node $i$, $\mathbf{q}_{i}$, and a {\it key} vector on the edge $ij$, $\mathbf{k}_{ij}$, given by, respectively,
\begin{align}
    \mathbf{q}_{i} &= \bigoplus_{l \ge 0}\sum_{k \ge 0}\mathbf{W}_Q^{lk}\mathbf{f}_{\text{in},i}^k, \\
    \mathbf{k}_{ij} &= \bigoplus_{l \ge 0}\sum_{k \ge 0}\mathbf{W}_K^{lk}(\mathbf{r}_j - \mathbf{r}_i)\mathbf{f}_{\text{in},j}^k.
\end{align}
$\bigoplus_{l \ge 0}$ denotes the direct sum over the $l$ channels.
Fig. \ref{fig: attention} depicts the message passing with the attention mechanism of the SE(3)-Transformer model.
Detailed numerical implementation of these models can be found in \url{https://github.com/trachote/predict\_elastic\_tensor}. 

\section{Macroscopic elastic moduli}
\setcounter{figure}{0}
Here are the formulae of elastic moduli computable from the elastic constants.   

\noindent Voigt's bulk modulus:
\begin{equation}\label{eq:bulk}
    B_V = \frac{1}{9}\sum_{i,j=1}^3C_{ij}.
\end{equation}
Voigt's shear modulus:
\begin{equation}\label{eq:shear}
    G_V = \frac{1}{30}\sum_{i,j=1}^3(C_{ii} - C_{ij}) + \frac{1}{5}\sum_{i=4}^6C_{ii}.
\end{equation}
Young's modulus:
\begin{equation}\label{eq:young}
    Y = \frac{9BG}{3B + G}.
\end{equation}
The Possion's ratio:
\begin{equation}\label{eq:possion}
    \nu = \frac{3B - 2G}{6B + 2G}.
\end{equation}

\section{Symmetry in the strain energy tensor}\label{sec: symmetry_SED}
\setcounter{figure}{0}
The strain energy density (SED) tensor ($\mathbf{U}$) is derived from the elastic tensor ($\mathbf{C}$). The symmetry in $\mathbf{U}$ follows the symmetry in $\mathbf{C}$ which depends on the crystal system.

\subsection{Hexagonal and Tetragonal I \\ (for Laue class $4/mmm$)}
\begin{equation}
    \mathbf{C}_{\textrm{hex/tet}} = 
    \begin{bmatrix}
    C_{11}&C_{12}&C_{13}&0&0&0\\[3pt]
     &C_{11}&C_{13}&0&0&0\\[3pt]
     & &C_{33}&0&0&0\\[3pt]
     & & &C_{44}&0&0\\[3pt]
     & & & &C_{44}&0\\[3pt]
     & & & & &C_{66}
    \end{bmatrix}
    .
\end{equation}
\begin{equation}
    \mathbf{U}_{\textrm{hex/tet}} = 
    \begin{bmatrix}
    U_{11}&U_{12}&U_{13}&U_{14}&U_{14}&U_{16}\\[3pt]
     &U_{11}&U_{13}&U_{14}&U_{14}&U_{16}\\[3pt]
     & &U_{33}&U_{34}&U_{34}&U_{36}\\[3pt]
     & & &U_{44}&U_{45}&U_{46}\\[3pt]
     & & & &U_{44}&U_{46}\\[3pt]
     & & & & &U_{66}
    \end{bmatrix}
    .
\end{equation}

\subsection{Tetragonal II (for Laue class $4/m$)}
\begin{equation}
    \mathbf{C}_{\textrm{tetII}} = 
    \begin{bmatrix}
    C_{11}&C_{12}&C_{13}&0&0&C_{16}\\[3pt]
     &C_{11}&C_{13}&0&0&-C_{16}\\[3pt]
     & &C_{33}&0&0&0\\[3pt]
     & & &C_{44}&0&0\\[3pt]
     & & & &C_{44}&0\\[3pt]
     & & & & &C_{66}
    \end{bmatrix}
    .
\end{equation}
\begin{equation}
    \mathbf{U}_{\textrm{tetII}} = 
    \begin{bmatrix}
    U_{11}&U_{12}&U_{13}&U_{14}&U_{14}&U_{16}\\[3pt]
     &U_{11}&U_{13}&U_{14}&U_{14}&U_{16}^{'}\\[3pt]
     & &U_{33}&U_{34}&U_{34}&U_{36}\\[3pt]
     & & &U_{44}&U_{45}&U_{46}\\[3pt]
     & & & &U_{44}&U_{46}\\[3pt]
     & & & & &U_{66}
    \end{bmatrix}
    ,
\end{equation}
where $U_{16}^{'}=-C_{16}\epsilon_1\epsilon_6 + \frac{1}{2}(C_{11}\epsilon_1^2 + C_{66}\epsilon_6^2)$.

\subsection{Rhombohedral I (for Laue class $\bar{3}m$)}
\begin{equation}
    \mathbf{C}_{\textrm{rhomI}} = 
    \begin{bmatrix}
    C_{11}&C_{12}&C_{13}&C_{14}&0&0\\[3pt]
     &C_{11}&C_{13}&-C_{14}&0&0\\[3pt]
     & &C_{33}&0&0&0\\[3pt]
     & & &C_{44}&0&0\\[3pt]
     & & & &C_{44}&C_{14}\\[3pt]
     & & & & &C_{66}
    \end{bmatrix}
    .
\end{equation}
\begin{equation}
    \mathbf{U}_{\textrm{rhomI}} = 
    \begin{bmatrix}
    U_{11}&U_{12}&U_{13}&U_{14}&U_{14}^{''}&U_{16}\\[3pt]
     &U_{11}&U_{13}&U_{14}^{'}&U_{14}^{''}&U_{16}\\[3pt]
     & &U_{33}&U_{34}&U_{34}&U_{36}\\[3pt]
     & & &U_{44}&U_{45}&U_{46}\\[3pt]
     & & & &U_{44}&U_{46}^{'}\\[3pt]
     & & & & &U_{66}
    \end{bmatrix}
    ,
\end{equation}
where $U_{14}^{'} = -C_{14}\epsilon_1\epsilon_4 + \frac{1}{2}(C_{11}\epsilon_1^2 + C_{44}\epsilon_4^2)$, $U_{14}^{''} = \frac{1}{2}(C_{11}\epsilon_1^2 + C_{44}\epsilon_4^2)$, $U_{46}^{'} = C_{14}\epsilon_4\epsilon_6 + \frac{1}{2}(C_{44}\epsilon_4^2 + C_{66}\epsilon_6^2)$. 
\newline 
\subsection{Rhombohedral II (for Laue class $3$)}
\begin{equation}
    \mathbf{C}_{\textrm{rhomII}} = 
    \begin{bmatrix}
    C_{11}&C_{12}&C_{13}&C_{14}&C_{15}&0\\[3pt]
     &C_{11}&C_{13}&-C_{14}&-C_{15}&0\\[3pt]
     & &C_{33}&0&0&0\\[3pt]
     & & &C_{44}&0&-C_{15}\\[3pt]
     & & & &C_{44}&C_{14}\\[3pt]
     & & & & &C_{66}
    \end{bmatrix}
    .
\end{equation}
\begin{equation}
    \mathbf{U}_{\textrm{rhomII}} = 
    \begin{bmatrix}
    U_{11}&U_{12}&U_{13}&U_{14}&U_{15}&U_{16}\\[3pt]
     &U_{11}&U_{13}&U_{14}^{'}&U_{15}^{'}&U_{16}\\[3pt]
     & &U_{33}&U_{34}&U_{34}&U_{36}\\[3pt]
     & & &U_{44}&U_{45}&U_{46}^{''}\\[3pt]
     & & & &U_{44}&U_{46}^{'}\\[3pt]
     & & & & &U_{66}
    \end{bmatrix}
    ,
\end{equation}
where $U_{46}^{''} = -C_{15}\epsilon_4\epsilon_6 + \frac{1}{2}(C_{44}\epsilon_4^2 + C_{66}\epsilon_6^2)$.
\newline 

\subsection{Orthorhombic, Monoclinic, and Triclinic}
\setcounter{figure}{0}
Strain energy tensors of orthorhombic, monoclinic, and triclinic lattices do not possess any degenerate structure, and are parametrized by the most general tensor of the form ~\eqref{eq:u_general} in the main text.

\section{The minimal set of strained components for each crystal systems in a training data set}
\setcounter{figure}{0}
The equivariant property of the SE(3)-transformer should be able to identify strained crystal graph inputs that are equivalent up to a rotation, thus to reduce the number of training data, we require one independent $ij$ component of the SED tensor for each degeneracy class vector. The other strained components can be incorporated during the randomized update of the one-hot strained component vector within the same degeneracy class in each training epoch, see Sec. \ref{subsec: model and training}. The minimal set of $ij$ indices of each crystal system can be chosen from the degenerate structure of the SED tensor for each crystal system in the previous Appendix \ref{sec: symmetry_SED}. Our chosen set is presented in Table~\ref{tab:crystal_system}.

\section{Dataset}\label{sec: dataset}
\setcounter{figure}{0}
Material data is pooled from the Materials Project database \cite{Jain2013}. There are 12,105 materials in the database that have the elastic tensor. However, only some materials are used in this work due to memory restriction. The memory usage increases with the number of edges, so the computationally feasible data is restricted to the materials with the maximum number of edges of 2,000 constructed by a cell-radii approach. There are 9,296 out of 12,105 materials from the database with this property (about 76.79\%). Fig.~\ref{fig:edge_size} shows that the selected materials have bulk and shear moduli distributing in the range of $0-600$ GPa, which can be a good representative of the whole database. In addition, we have found that the validation errors do not decrease significantly even when the number of edges increases to 3,000. The statistics of crystal systems curated for training and test sets of these 9,296 materials is shown is Table~\ref{tab:crystal_system_data}. The dataset is partitioned into a training set and a validation set using the stratified method (in scikit-learn) with a random seed number 0.

\begin{figure*}[h!]
\includegraphics[width=0.75\textwidth]{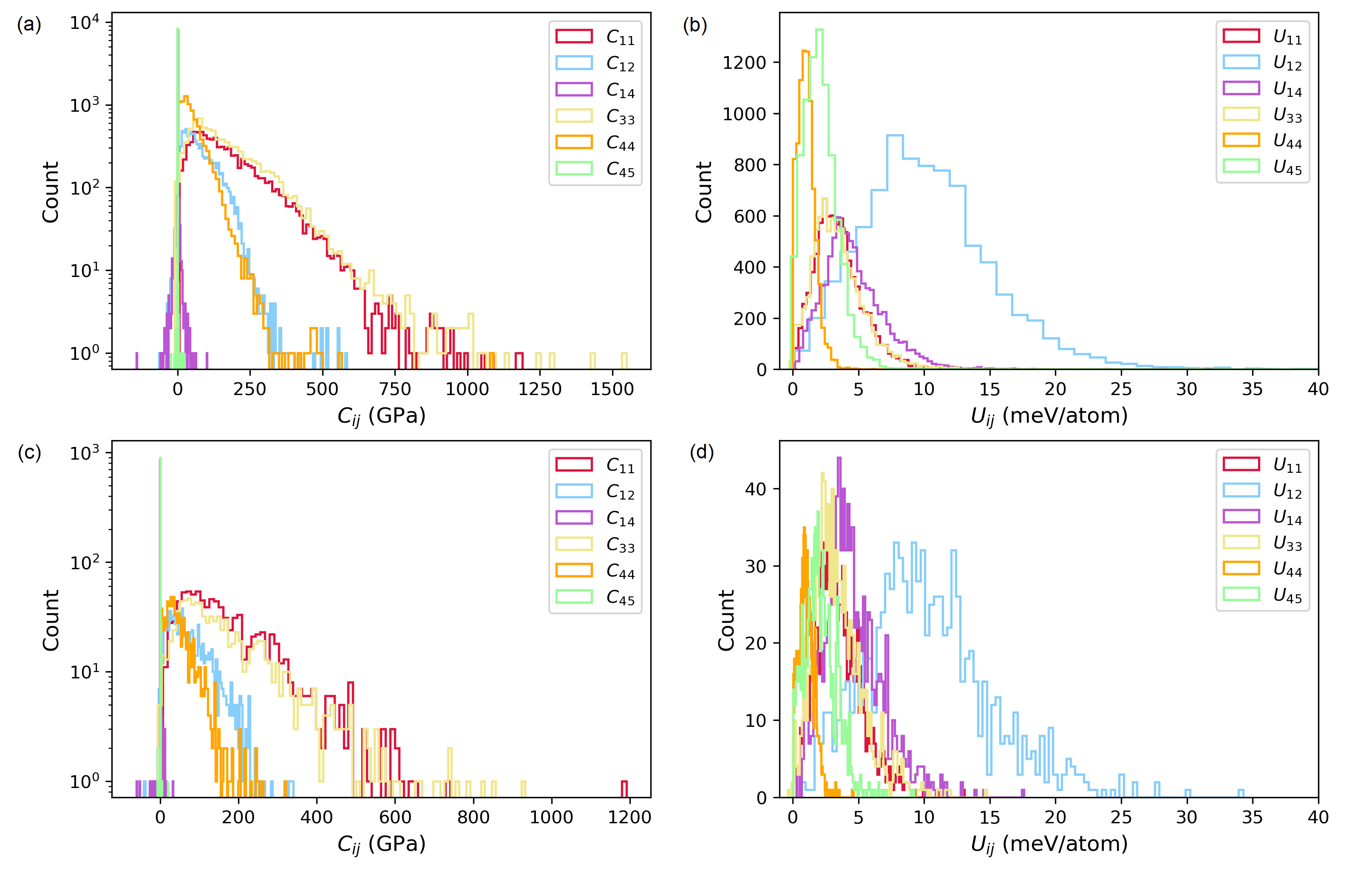}
\caption{\label{fig:data_distributions}
The histograms of (a) $C_{ij}$ and of (b) $U_{ij}$ in the training set and of (c) $C_{ij}$ and  of (d) $U_{ij}$ in the test set. For visualization purposes, only some $ij$ components, specifically 11, 12, 14, 33, 44, and 45, are presented. For each $ij$ component, $U_{ij}$ ($C_{ij}$) are distributed over different range of energy (modulus). For $i\neq j$, $U_{ij}$ are distributed over a broader range than $U_{ii}$ are. For $ij=45$, the elastic constant is concentrated around 0 GPa. Importantly, this concentration around zero issue can be ameliorated when the elastic constants are converted to the strain energy density. This conversion to SED reduces the number of zero values in the dataset, which significantly improves model training and model predictive performances.}
\end{figure*}

\begin{figure*}[h]
\includegraphics[width=0.5\textwidth]{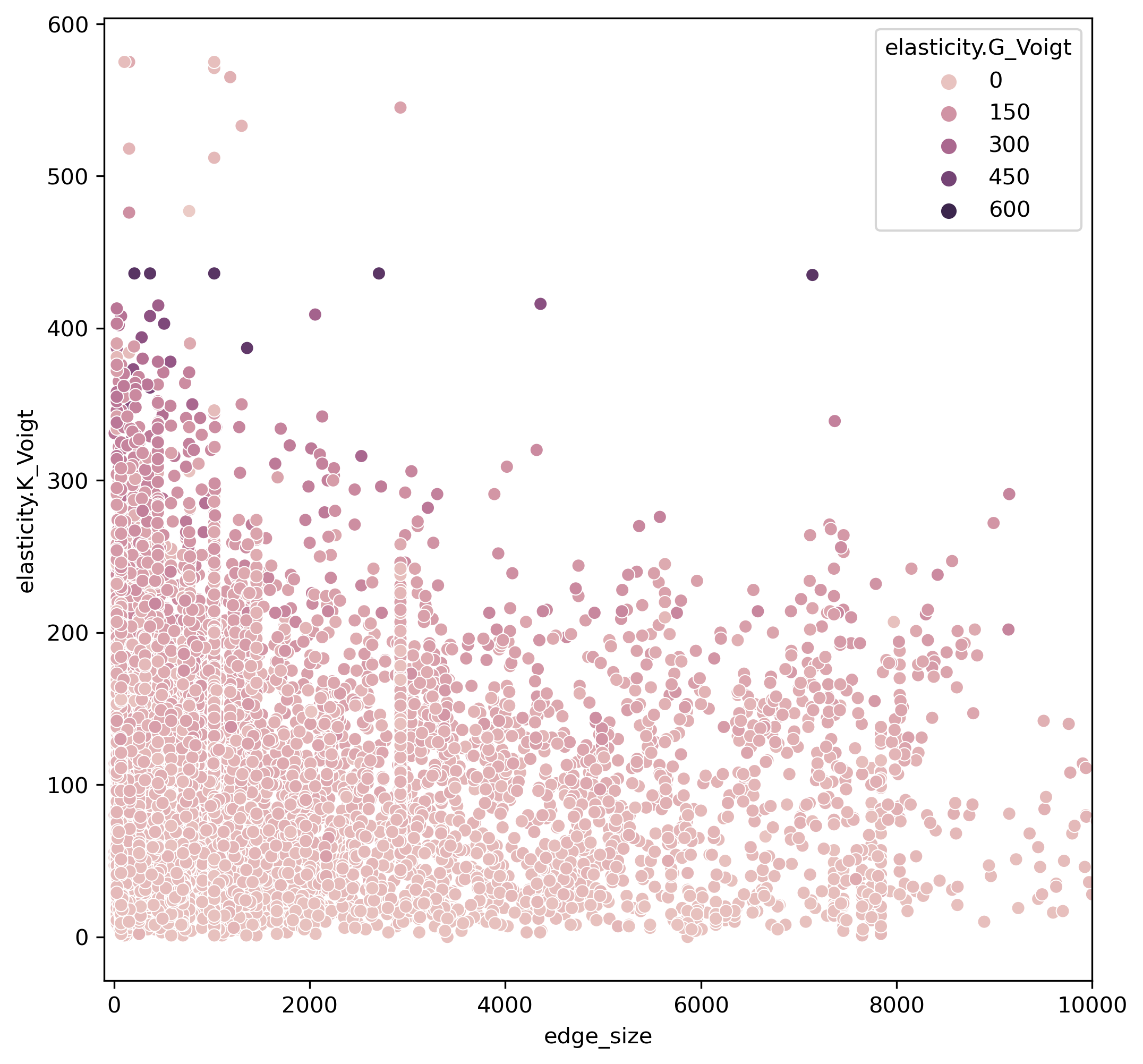}
\caption{\label{fig:edge_size}The bulk modulus of materials is plotted against the number of edges created by cell radii method, with the hue indicating the shear modulus. The data below a hard cutoff of 2000 edges can already cover the range of values of bulk and shear moduli of the whole dataset.}
\end{figure*}

\begin{table}[h]
\caption{\label{tab:edge_size}Statistics of number of edges for 12105 materials created by cell radii and CrystalNN methods.}
\begin{ruledtabular}
\begin{tabular}{lcc}
&Cell radii&CrystalNN\\
\hline
mean&2115.16&143.47\\
std&4665.66&129.67\\
min&6&6\\
25\%&368&58\\
50\%&1024&112\\
75\%&1776&224\\
max&75864&2272
\end{tabular}
\end{ruledtabular}
\end{table}

\begin{table}[h]
\caption{\label{tab:crystal_system_data}The number of data for training and test sets categorized by crystal systems.}
\begin{ruledtabular}
\begin{tabular}{lcccc}
\multicolumn{1}{c}{Crystal system} & \multicolumn{2}{c}{Training set} & \multicolumn{2}{c}{Test set} \\
& Count & \% & Count & \% \\
\hline
Cubic & 3865 & 46.17 & 441 & 47.83\\
Hexagonal & 1369 & 16.35 & 146 & 15.84\\
Tetragonal & 1580 & 18.87 & 163 & 17.68\\
Trigonal & 530 & 6.33 & 60 & 6.51\\
Orthorhombic & 731 & 8.73 & 79 & 8.57\\
Monoclinic & 253 & 3.02 & 27 & 2.93\\
Triclinic & 43 & 0.51 & 6 & 0.65
\end{tabular}
\end{ruledtabular}
\end{table}

\section{Model parameters}
The hyperparameters that significantly affect the model performance are the number of convolution layers, the number of $l$-channels in the SE(3) kernel, and the number of hidden units of the learnable radial function in the kernel $\phi$. The number of convolution layers is 4, where the model performance only slightly improves with more layers.  The learnable radial function $\phi$ is a feedforward network, consisting of an input layer that is fed into a fully-connected hidden layer with a ReLU activation, which is forwarded to another fully-connected layer that outputs a hidden feature vector.  We use the number of hidden units in the hidden layer of $\phi$ to be 128 to avoid over-squashing, and the performance only marginally improves expanding beyond 128 hidden units. The dimension of the hidden feature vectors  $\mathbf{f}^l_{\mathrm{hid},n}$ is 32, whereas the dimension of $\mathbf{f}^l_{\mathrm{reg},n}$ and $\mathbf{f}^l_{\mathrm{class},n}$ are 128. The dimension of the  input atomic number embedding is 512. The maximum degree of $l$ is 4 where, of course, the model performance also improves if the degree is higher especially for low-symmetry crystal structures; however, increasing the maximum degree of $l$ beyond 4 becomes infeasible on our computational infrastructure, with other hyperparameters fixed. The number of transformer-head is 2. We use ADAM optimizer and cosine annealing method for a scheduler with the initial learning rate of $10^{-4}$.

\section{Trial models and additional results}
\setcounter{figure}{0}

\begin{table*}[h!]
\caption{\label{tab:model_c}The MAE and RMSE of predicted elastic properties using the model in Fig.~\ref{fig:additional_models}(b). The dataset for this model consists of materials whose number of edges is less than 2000.}
\begin{ruledtabular}
\begin{tabular}{cccc}
Properties & Average & MAE & RMSE\\
\hline
$U_{ij}$ (meV/atom) & 2.652 & 0.661 & 1.258\\
$C_{ij}$ (GPa) & 42.92 & 10.98 & 17.86\\
$B_V$ (GPa) & 107.07 & 11.58 & 19.10 \\
$G_V$ (GPa) & 50.98 & 9.91 & 16.31 \\
$Y$ & 129.62 & 22.81 & 37.24 \\
$\nu$ & 0.401 & 0.035 & 0.081 
\end{tabular}
\end{ruledtabular}
\end{table*}

\begin{figure*}[h!]
\includegraphics[width=.90\textwidth]{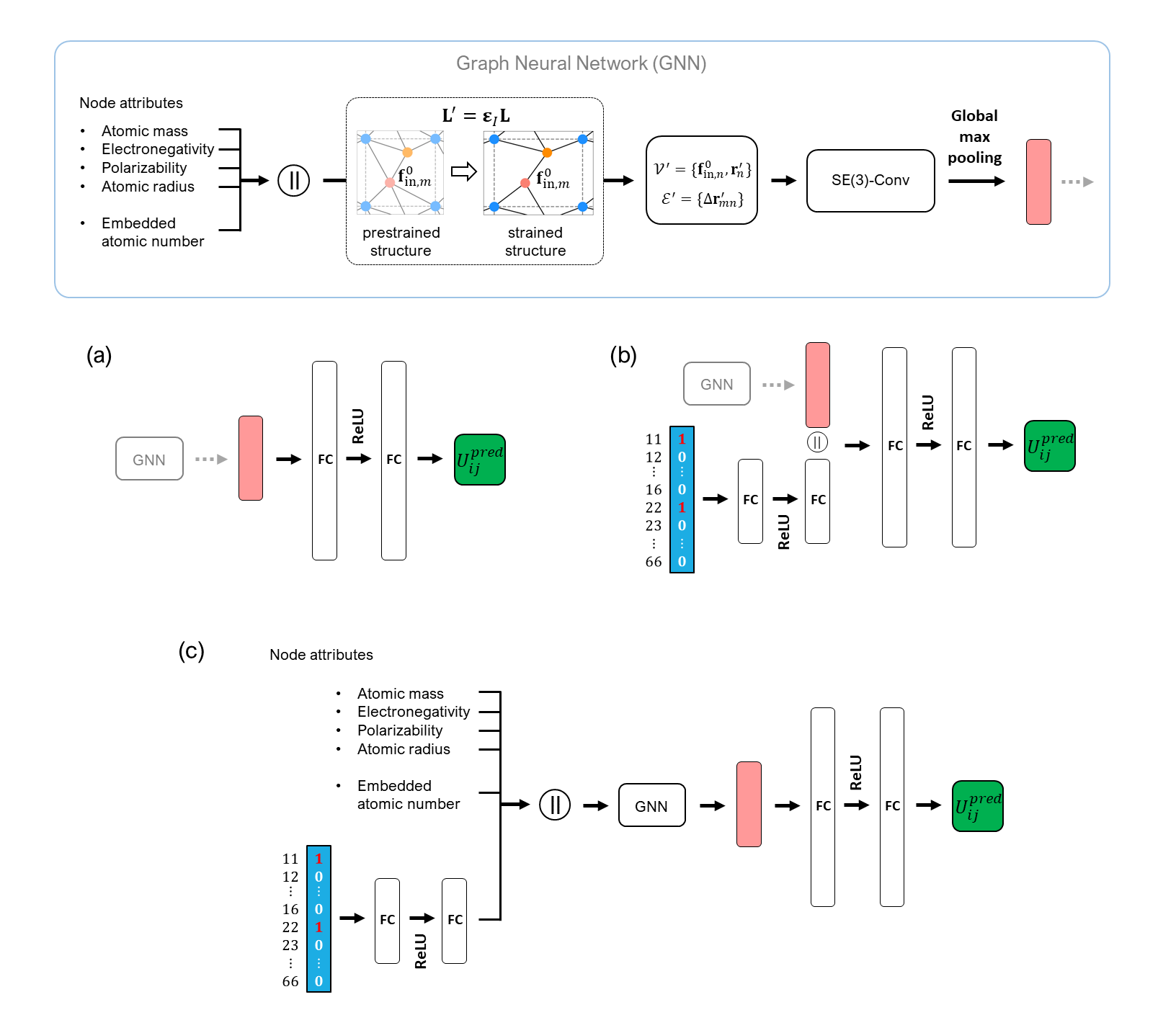}
\caption{\label{fig:additional_models} 
Earlier trial models with SE(3)-equivariant GNNs as a fundamental building block. (Top) The building block consists of a strained crystal graph input that is fed into SE(3)-equivariant GNNs (SE(3)-transformer followed by a TFN, similar to that of the building block in \textcolor{black}{StrainTensorNet}), whose output is the global max-pooled feature from the TFN.  (a) A minimal equivariant GNN model for SED regression. (b) An SED  regression network that take both the known degeneracy class vector and the minimal equivariant GNN's representation as an input. (c) The same network as (a) but the input node attributes also incorporates the degeneracy class vector.}
\end{figure*}

\begin{figure*}[h!]
\includegraphics[width=1.\textwidth]{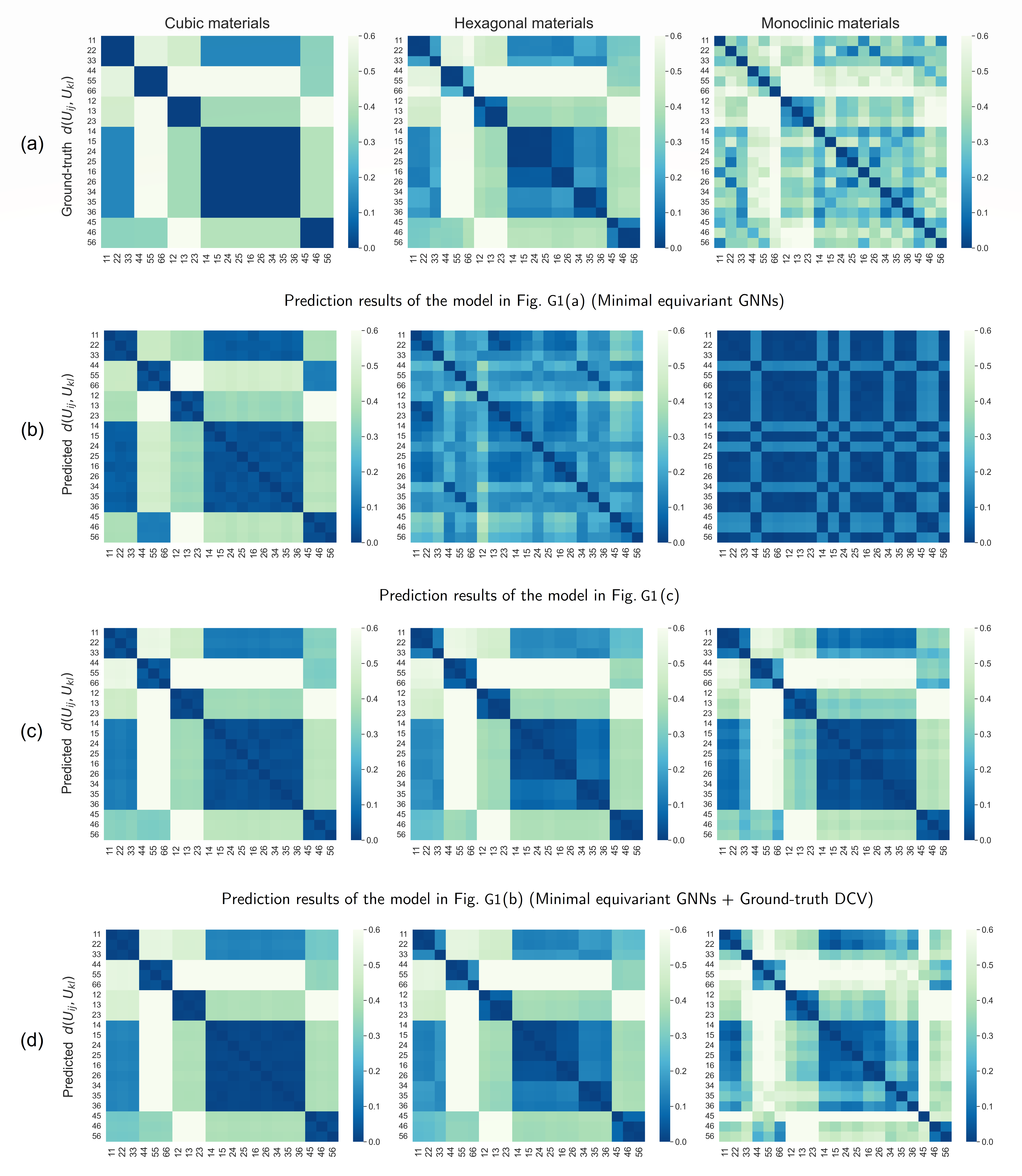}
\caption{\label{fig:additional_dm} 
From top to bottom rows are, respectively, the distance matric (of the materials whose numbers of edges are less than 500) of the ground truths, of the model in Fig.~\ref{fig:additional_models}(a), of the model Fig.~\ref{fig:additional_models}(c), and of the model model in Fig.~\ref{fig:additional_models}(b). The last row suggests that the concatenation of the strained crystal graph latent representation and the degeneracy class vector significantly increases the expressiveness of the SED regression network, motivating the self-supervised network architecture in the main text.
}
\end{figure*}

\begin{table*}
\caption{\label{tab:models_bcd}The MAE of predicted elastic properties using the model in Fig.~\ref{fig:additional_models}(a), (b), and (c). The dataset for these earlier models are the materials whose number of edges is less than 500.}
\begin{ruledtabular}
\begin{tabular}{ccccc}
Properties & Average & MAE of Fig.~\ref{fig:additional_models}(a) & MAE of Fig.~\ref{fig:additional_models}(b) & MAE of Fig.~\ref{fig:additional_models}(c)\\
\hline
$U_{ij}$ (meV/atom) & 2.749 & 1.628 & {\bf 0.842} & 0.866 \\
$C_{ij}$ (GPa) & 44.10 & 78.90 &  {\bf 13.29} & 18.66 \\
$B_V$ (GPa) & 108.67 & 67.95 &  {\bf 15.35} & 17.07 \\
$G_V$ (GPa) & 53.92 & 79.94 &  {\bf 12.46} & 14.63 \\
$Y$ & 136.26 & 86.05 &  {\bf 29.08} & 31.93 \\
$\nu$ & 0.394 & 1.758 &  {\bf 0.044} & 0.061
\end{tabular}
\end{ruledtabular}
\end{table*}


\begin{figure*}[h]
\includegraphics[width=0.7\textwidth]{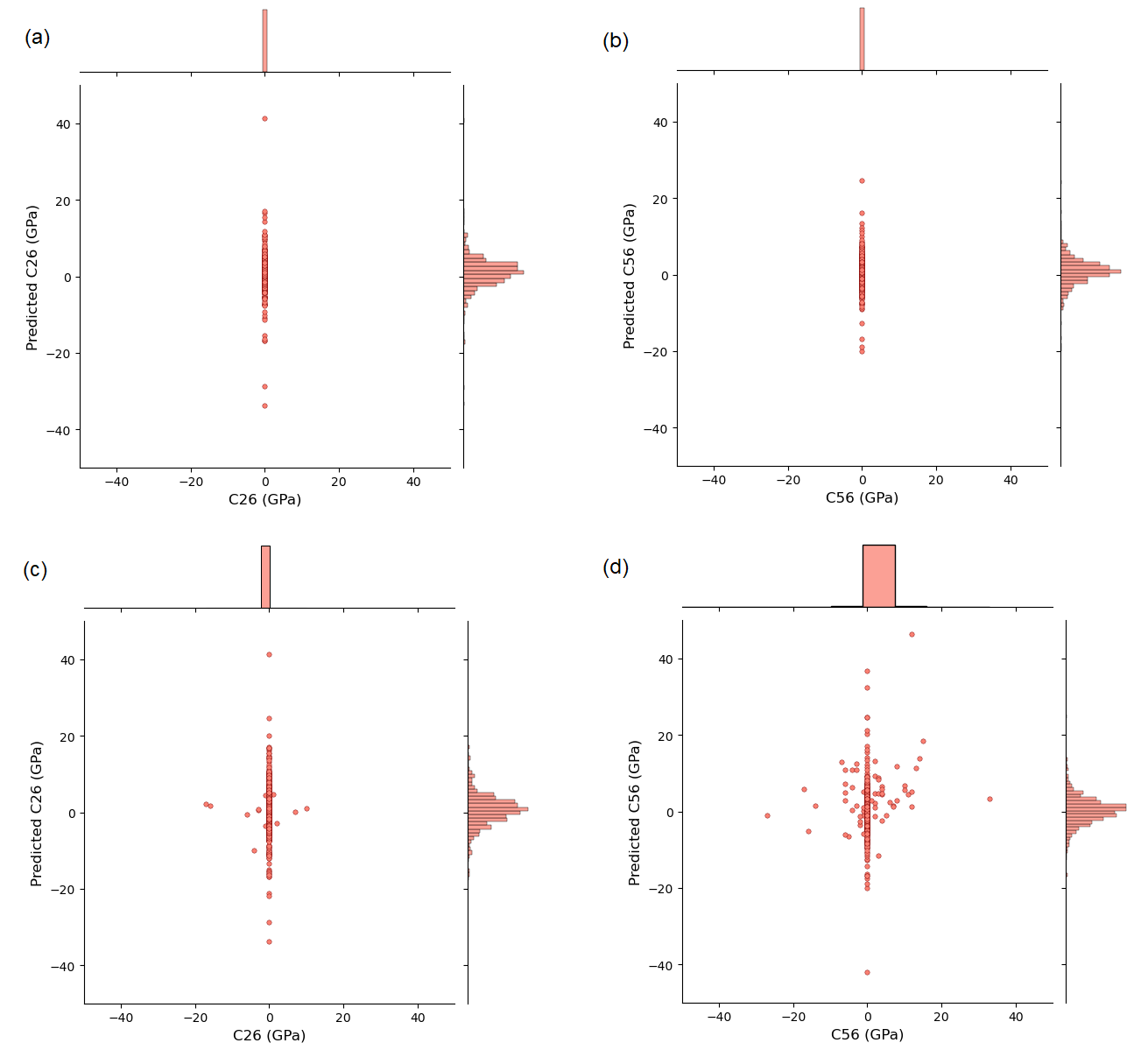} 
\caption{\label{fig:jointplots} Statistics of the prediction results (vertical axes) vs of the ground truths (horizontal axes) of the elastic constant components that are concentrated near 0 GPa. (a)  The predicted $C_{26}$ of a cubic system has the mean and the standard deviation of 0.69 and 4.45 GPa. (b) The predicted  $C_{56}$ of a cubic system has the the mean and the standard deviation of -0.43 and 4.85 GPa. (c) The predicted  $C_{26}$ of all the crystal systems has the mean and the standard deviation of 0.99 and 6.08 GPa. (d) The predicted  $C_{56}$ of all the crystal systems has the mean and the standard deviation of -0.20 and 9.22 GPa, respectively.  The ground-truth distributions of $C_{26}$ and $C_{56}$ are centered around 0 GPa with vanishing widths. Although some of the predicted $C_{26}$ and $C_{56}$ are not exactly 0 GPa (for cubic system), most of the predicted values are near 0 GPa. }
\end{figure*}

\begin{figure*}[h]
\includegraphics[width=1.0\textwidth]{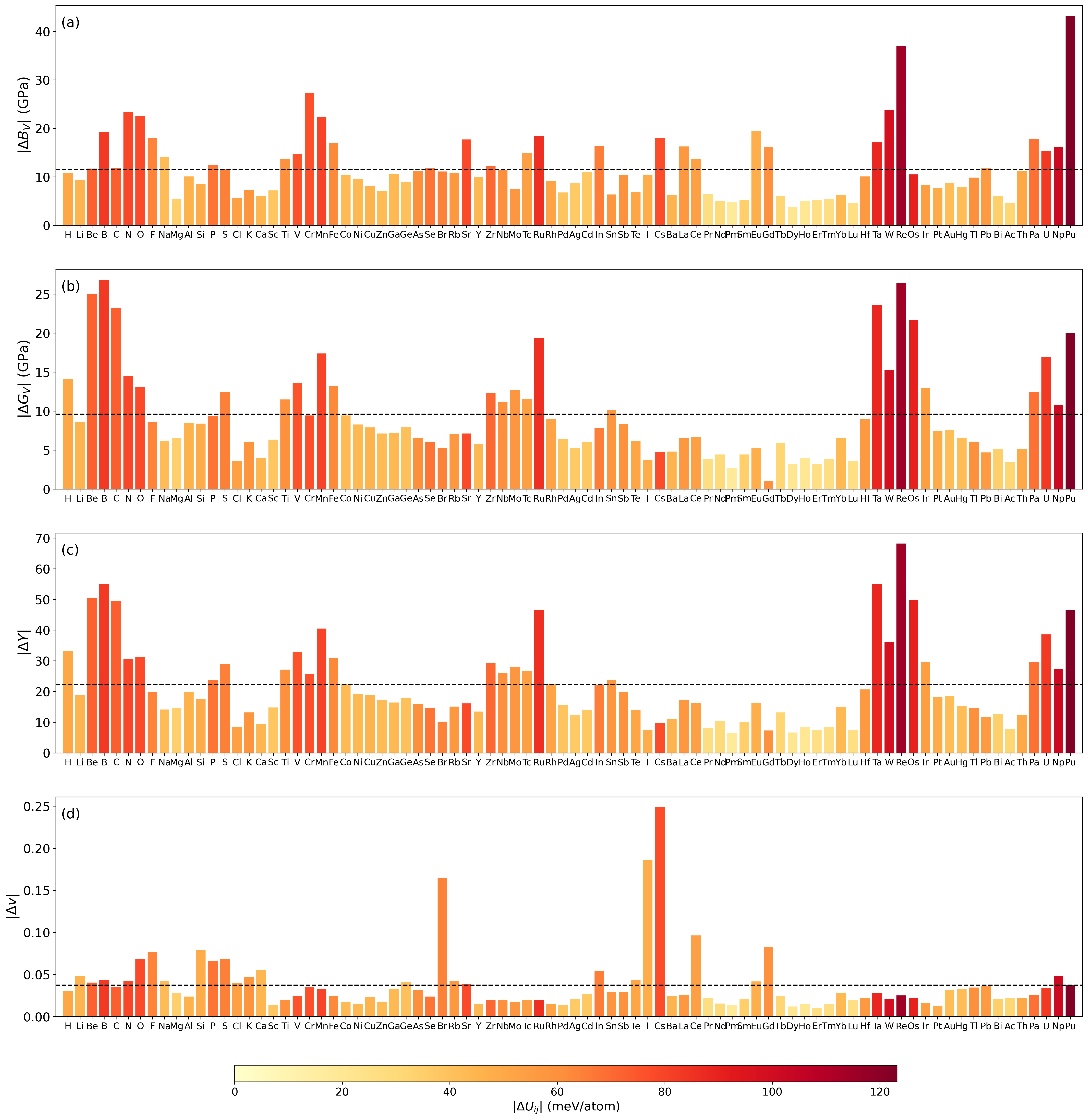}
\caption{\label{fig:errors_probs} 
The bar charts display the MAE of the predicted elastic properties, i.e., (a) $B_V$, (b) $G_V$, (c) $Y$, and (d) $\nu$. The charts are categorized based on the elements present in the compounds in the test set and colored according to the average bulk modulus, as indicated in the legend below. The dashed lines represent the MAE of each property, averaged over the data in the test set. It can be seen that most elements, whose compounds have high bulk modulus, exhibit errors of $B_V$, $G_V$, and $Y$ that exceed the dataset's MAE.
}
\end{figure*}

\end{appendix}

\nocite{*}

\bibliography{references}

\end{document}